\documentclass[nofootinbib,eqsecnum,superscriptaddress,amsmath,amssymb,aps,pra,11pt]{revtex4-2}

\usepackage{amsfonts}
\usepackage{bbm} 
\usepackage{mathrsfs} 
\usepackage{multirow}

\usepackage{stackengine} 

\usepackage{graphicx}

\usepackage{footnotebackref} 

\DeclareMathOperator{\Tr}{Tr} 

\newcommand{\ket}[1]{\vert{#1}\rangle} 
\newcommand{\bra}[1]{\langle{#1}\vert} 

\newcommand{\secref}[1]{Sec.~\ref{#1}}

\newcommand{\eqnref}[1]{(\ref{#1})}
\newcommand{\figref}[1]{Fig.~\ref{#1}}
\newcommand{\appref}[1]{Appendix~\ref{#1}}
\newcommand{\ftref}[1]{Footnote~\ref{#1}} 

\begin{document}
\count\footins = 1000 

\title{Complementarity relations of a delayed-choice quantum eraser in a quantum circuit}


\author{Dah-Wei Chiou}
\email{dwchiou@gmail.com}
\thanks{}
\affiliation{Department of Electrical Engineering, National Taiwan University, Taipei 10617, Taiwan}

\author{Hsiu-Chuan Hsu}
\email{hcjhsu@nccu.edu.tw}
\thanks{The two authors contributed equally.}
\affiliation{Graduate Institute of Applied Physics, National Chengchi University, Taipei 11605, Taiwan}
\affiliation{Department of Computer Science, National Chengchi University, Taipei 11605, Taiwan}


\begin{abstract}
We propose a quantum circuit that emulates a delayed-choice quantum eraser via bipartite entanglement with the extension that the degree of entanglement between the two paired quantons is adjustable. This provides a broader setting to test complementarity relations between interference visibility and which-way distinguishability in the scenario that the which-way information is obtained through entanglement without direct contact with the quantum state for interference. The visibility-distinguishability relations are investigated from three perspectives that differ in how the which-way information is taken into consideration. These complementarity relations can be understood in terms of entropic uncertainty relations in the information-theoretic framework and the triality relation that incorporates single-particle and bipartite properties. We then perform experiments on the quantum computers provided by the IBM Quantum platform to verify the theoretical predictions. We also apply the delay gate to delay the measurement of the which-way information to affirm that the measurement can be made truly in the ``delayed-choice'' manner.
\end{abstract}

\maketitle


\section{Introduction}
Wave-particle duality is a fundamental concept of quantum mechanics that is closely related to Bohr's complementarity principle \cite{bohr1928,scully1991quantum}. It holds that every quantum object possesses both wave and particle properties, but the two properties cannot be observed or measured simultaneously. For single quantons (e.g., photons) in a two-path Mach-Zehnder interferometer, the wave behavior is quantified by interference visibility $\mathcal{V}$, which indicates how sharp the interference fringe pattern is, and the particle behavior is quantified by path distinguishability $\mathcal{D}$, which indicates how much the ``which-way'' information about which path the particle has travelled can be inferred. In terms of $\mathcal{V}$ and $\mathcal{D}$, an information-theoretic formulation developed by Jaeger \textit{at al.}\ \cite{PhysRevA.51.54} and Englert \cite{PhysRevLett.77.2154} leads to a wave-particle duality relation
\begin{equation}\label{wave-particle duality}
\mathcal{V}^2+\mathcal{D}^2\le1,
\end{equation}
which quantitatively bounds the complementarity between the wave and particle behaviors.

The similar analysis has been generalized for more complicated settings, such as in multipath interferometers \cite{PhysRevA.64.042113} and in the context of quantum erasure \cite{PhysRevA.58.3477,ENGLERT2000337}.
The duality relation for bipartite systems has also been extended to a ``triality'' relation, which additionally takes into account a quantitative measure of entanglement concurrence \cite{PhysRevA.76.052107,Jakob2010}.
The triality relation has also been generalized to multipath scenarios \cite{PhysRevA.105.032209}.

It has been debated whether the duality relation $\mathcal{V}^2+\mathcal{D}^2\leq1$ can be understood in terms of Heisenberg's uncertainty principle, which gives an uncertainty relation for any two noncommuting observables. Originally, it was argued that the two principles are independent \cite{PhysRevLett.77.2154}. Later, however, it was demonstrated that they are related to each other \cite{durr2000can,BUSCH20061,coles2014equivalence}.
The work of \cite{coles2014equivalence} developed a new information-theoretic formulation, which casts Heisenberg's uncertainty relations as entropic uncertainty relations and therefore unifies Heisenberg's uncertainty principle with the particle-wave duality. This entropic formulation for wave-particle duality has been further established for multipath interferometers \cite{PhysRevA.93.062111, PhysRevA.92.012118, QURESHI2017598}. Other wave-particle duality relations have also been introduced in terms of different information-theoretic quantifiers \cite{angelo2015wave,PhysRevLett.116.160406}.

There are two very different scenarios for deducing the which-way information. One is to directly probe the particle's passage, whereas the other obtains the which-way information through entanglement. The first scenario disturbs the particle's passage, and thus it is not surprising that the interference pattern is diminished if the which-way information is known. The second scenario, by contrast, does not have any direct contact with the particle's passage at all, and it is somewhat counterintuitive that the interference pattern nevertheless depends on the deduced which-way information.
A typical example of the second scenario is a delayed-choice quantum eraser (see \cite{RevModPhys.88.015005} for a review), wherein, in a sense, the interference visibility can be ``enhanced by erasing the which-way information retroactively''.
As the entropic framework proposed by \cite{PhysRevA.93.062111} is very generic and applicable to both scenarios, the visibility-distinguishability tradeoff in an entanglement quantum eraser should still satisfy the inequality \eqref{wave-particle duality}. However, there are some features peculiar to the second scenario that are not addressed in \cite{PhysRevA.93.062111} and have to be understood from the triality relation proposed by \cite{Jakob2010}. It will shed new light on the wave-particle duality to study the complementarity between $\mathcal{V}$ and $\mathcal{D}$ for entanglement quantum erasure. This paper aims to investigate this aspect.

The inequality \eqref{wave-particle duality} applies not only to interference of different paths in an interferometer but also interference of any alternative states. Typically, one can emulate a two-path interference experiment in a quantum circuit by considering the interference between the qubit states $\ket{0}$ and $\ket{1}$.
Reframing the delayed-choice experiment in a quantum circuit provides a higher level of abstraction that renders the information flow more transparent \cite{PhysRevLett.107.230406}.
Today, IBM Quantum provides an online platform that allows users to access the cloud services of quantum computing \cite{IBMQ}. It offers an accessible and easily manageable facility for performing interference experiments. The IBM Quantum platform has been used to carry out various interference experiments and study visibility-distinguishability duality relations \cite{PhysRevA.102.032605,PhysRevA.103.022409,PhysRevA.103.022212,PhysRevA.104.032223}.
Particularly, the recent work of \cite{PhysRevA.104.032223} studies the visibility-distinguishability complementarity for a quantum circuit, where the which-way information is deduced via a minimum-error measurement and via a nondelayed quantum eraser, both in the aforementioned first scenario.

In this paper, we consider a quantum circuit in analogy to a delayed-choice quantum eraser and investigate the visibility-distinguishability complementarity in the second scenario.
The quantum circuit of a delayed-choice quantum eraser not only is more easily implemented than an optical experiment, but it also has the merit that the degree of entanglement between the two paired quantons is adjustable, thus facilitating the analysis of a delayed-choice quantum eraser in a broader setting with full or partial entanglement, which cannot be easily implemented in an optical experiment.
The complementarity relation between visibility and distinguishability is then investigated in depth from three different perspectives: (i) in view of the total ensemble of all events, (ii) in view of separate subensembles associated with different readouts of the entangled qubit, and (iii) in view of average results averaged over subensembles.
The analysis provides valuable insight into the intricate interplay between visibility, distinguishability, and bipartite entanglement.

We then perform experiments on the quantum computers of IBM Quantum. The experimental results agree very well with the theory. The deviations from theory are insignificant except in some extreme cases where the error of the CNOT gate results in deviations that become considerable only in perspective (ii). Moreover, we utilize the delay gate that instructs a qubit to idle for a requested duration to fulfill a truly ``delayed-choice'' measurement. As a consequence of decoherence over time, which corrupts the entanglement between the two qubits, the interference pattern is ``recovered'' by the quantum eraser effect to a lesser extent compared to the nondelayed case, and various visibility and distinguishability quantifiers are also diminished to a certain degree.

This paper is organized as follows. In \secref{sec:quantum eraser}, we first propose a simple model of an entanglement quantum eraser in an optical experiment. In \secref{sec:quantum circuit}, we implement a quantum circuit in analogy to the entanglement quantum eraser with the extension that the degree of the bipartite entanglement is adjustable. In \secref{sec:V and D}, we calculate various quantifiers of visibility and distinguishability and investigate the complementarity relations of them from three different perspectives. In \secref{sec:comparison}, we compare these relations to those established in the literature. In \secref{sec:IBM Q}, we present and analyze the experimental results performed on IBM Quantum. Finally, the theoretical and experimental results are summarized in \secref{sec:summary}. Various technical details and supplementary data are provided in the appendices.

\section{A simple model of a delayed-choice quantum eraser}\label{sec:quantum eraser}
The idea of delayed-choice quantum erasure was first proposed by Scully and Dr{\"u}hl in 1982 \cite{PhysRevA.25.2208}. Since then, many different scenarios framing the same concept have been conceived. The first quantum eraser experiment was performed by Kim \textit{et al.}\ in 1999 \cite{PhysRevLett.84.1} in a double-slit interference experiment using entangled photons. A similar double-slit experiment involving photon polarization was later performed by Walborn \textit{et al.}\ in 2002 \cite{PhysRevA.65.033818}. (See \cite{RevModPhys.88.015005} for a comprehensive review.)

\begin{figure}
\centering
    \includegraphics[width=0.75\textwidth]{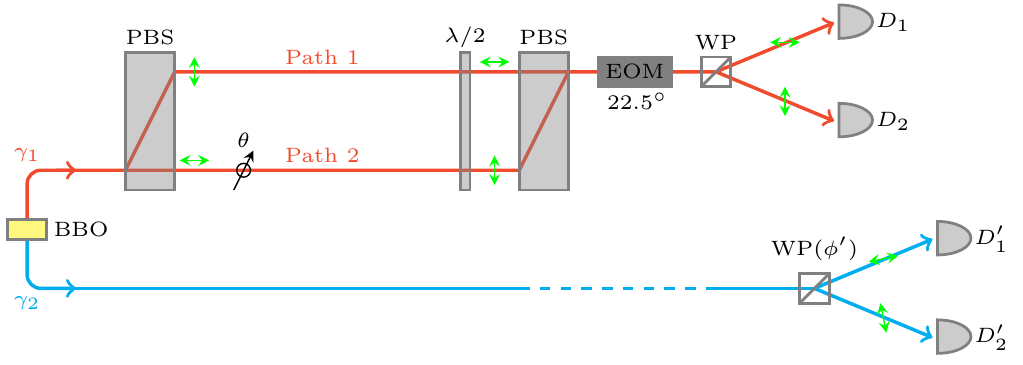}
\caption{Experimental setup for the delayed-choice quantum eraser. A pair of entangled photons $\gamma_1$ and $\gamma_2$ with orthogonal polarizations are created from a nonlinear optical crystal BBO. The photon $\gamma_1$ is directed into a Mach-Zehnder interferometer with the detectors $D_1$ and $D_2$. On the other hand, the entangled partner $\gamma_2$ is directed into the delayed-choice measuring device with the detectors $D'_1$ and $D'_2$.}
\label{fig:interferometer}
\end{figure}

To give a simple description of the quantum eraser, we reformulate the double-slit experiment of \cite{PhysRevA.65.033818} in terms of a Mach-Zehnder interferometer, which is conceptually more concise and draws a close analogy implementable in a quantum circuit.\footnote{The same idea of using a Mach-Zehnder interferometer for the delayed-choice quantum eraser has been considered in the literature (see e.g., \cite{Qureshi_2020,qureshi2021delayed}). Particularly, our setup in \figref{fig:interferometer} is very similar to Figure~1 in \cite{qureshi2021delayed}, except that the latter investigates a different issue and does not consider an adjustable phase shift.} As illustrated in \figref{fig:interferometer}, spontaneous parametric down-conversion (SPDC) in a nonlinear optical crystal, such as beta barium borate (BBO), is used to prepare a pair of entangled photons ($\gamma_1$ and $\gamma_2$) that are orthogonally polarized. The photon $\gamma_1$ is directed into a Mach-Zehnder interferometer with the detectors $D_1$ and $D_2$, while the entangled partner $\gamma_2$ is directed into the ``delayed-choice'' measuring device with the detectors $D'_1$ and $D'_2$. The Mach-Zehnder interferometer is based on the setup of \cite{jacques2007experimental}, which was originally designed to realize Wheeler's delayed-choice experiment. Initially, the pathway of $\gamma_1$ is split by a polarizing beam splitter (PBS) into two spatially separated paths (Path 1 and Path 2) associated with vertical and horizontal polarizations, respectively. An adjustable phase-shift plate is inserted to Path 2 to provide a relative phase shift $\theta$ between the two paths. The two paths pass through a half-wave plate that rotates the photon polarization by $90^\circ$, and are recombined by a second polarizing beam splitter.\footnote{The adjustable phase shift $\theta$ can also be realized by tilting the second beam splitter.} The recombined path then enters an electro-optical modulator (EOM) of which the optical axis is oriented at $22.5^\circ$ from the direction of input polarization. Applied with the half-wave voltage ($V_\pi$), the EOM behaves as a half-wave plate that rotates the input photon polarization by $45^\circ$. Finally, a Wollaston prism is used to deflect horizontally polarized photons to $D_1$ and vertically polarized photons to $D_2$. When the voltage $V_\pi$ is applied, the whole interferometer functions as the ``closed'' configuration of the Wheeler's delayed-choice experiment (i.e., the two paths are recombined). On the other hand, if no voltage is applied, it effectively functions as the ``open'' configuration (i.e., the two paths do not interfere at all; Path 1 and Path 2 arrive at $D_1$, and $D_2$ separately).\footnote{The main merit of using the EOM is that the switch between the closed and open configurations can be made very fast, which is crucial for Wheeler's delayed-choice experiment. If the switch speed is not a concern, one can simply replace the EOM with a half-wave plate that rotates the input polarization by $45^\circ$ for the closed configurations, and remove the half-wave plate for the open configuration. The arrangement devised to recombine the two paths can alternatively be replaced by the method proposed in Figure~1 in \cite{qureshi2021delayed}.} The closed configuration is used for the quantum eraser experiment.

If we shed the single photon $\gamma_1$ into the interferometer and repeat the experiment many times, we obtain the detection probabilities of $D_1$ and $D_2$ from the accumulated counts of individual signals.
Because $\gamma_1$ and $\gamma_2$ are maximally entangled, $\gamma_1$ itself is completely unpolarized and thus its polarization is described by the density matrix $\rho_{\gamma_1}=\frac{1}{2}\mathbbm{1}_{2\times2}$. As $\rho_{\gamma_1}$ can be interpreted as having either horizontal or vertical polarization by a fifty-fifty chance, for each individual event of the accumulated ensemble, the photon $\gamma_1$ can be said to travel \emph{either} Path 1 \emph{or} Path 2 with equal probability. As a result of the interpretation that Path 1 and Path 2 are travelled separately, the detection probabilities of $D_1$ and $D_2$ are 50\% for each, showing no interference between the two paths (i.e., independent of the relative phase shift $\theta$).\footnote{\label{foot:unitary freedom}However, because of the \emph{unitary freedom} for density matrices, $\rho_{\gamma_1}$ also admits infinitely many different interpretations. For example, $\frac{1}{2}\mathbbm{1}_{2\times2}$ can be alternatively interpreted as having an arbitrarily specific polarization (e.g., clockwise polarized) by 50\% probability and having the orthogonal polarization (e.g, counterclockwise polarized) by the other 50\%. The resulting detection probabilities of $D_1$ and $D_2$ nevertheless are independent of the interpretation. If a different interpretation is adopted, each individual photon $\gamma_1$ may not be said to travel \emph{either} of the two paths and thus the two paths still interference to a certain degree depending on the interpretation, but the probabilistic nature of $\frac{1}{2}\mathbbm{1}_{2\times2}$ turns out to ``conceal'' the two-path interference of each individual event in the accumulated result.}

Meanwhile, as shown in the lower part of \figref{fig:interferometer}, the entangled partner $\gamma_2$ is directed into a Wollaston prism that splits two mutually orthogonal polarizations into the detectors $D'_1$ and $D'_2$ separately. The orientation of the Wollaston prism can be adjusted by an angle $\phi'$ so that the linear polarization at the angle $\phi'$ from the horizontal direction enters $D'_1$ while the linear polarization at the angle $\pi/2+\phi'$ enters $D'_2$. The polarization of $\gamma_2$ can be determined by whether it registers a signal at $D'_1$ or $D'_2$. Because $\gamma_1$ and $\gamma_2$ are entangled in the way that their polarizations are orthogonal to each other, the polarization of $\gamma_1$ can be inferred from knowing the polarization of $\gamma_2$. If we adjust $\phi'=0$, the events that click $D'_1$ correspond to horizontal polarization, and those that click $D'_2$ correspond to vertical polarization. The accumulated events of $\gamma_1$ measured by $D_1$ and $D_2$ can be grouped into two subensembles according to whether $\gamma_2$ clicks $D'_1$ or $D'_2$. Each individual event of $\gamma_1$ in the subensembles associated with $D'_1$ and $D'_2$ is said to travel along Path 1 and Path 2, respectively. As the ``which-way'' information of whether each individual $\gamma_1$ travels along Path 1 or Path2 has been ``marked'' by the associated $D'_1$ or $D'_2$ outcome, within each subensemble the detection probabilities of $D_1$ and $D_2$ remains 50\% for each, showing no two-path interference.

On the other hand, if we adjust $\phi'=\pi/4$, the events that click $D'_1$ correspond to the $45^\circ$ diagonal polarization, and those that click $D'_2$ correspond to the $135^\circ$ diagonal polarization. In the subensembles associated with $D'_1$ and $D'_2$, the photon $\gamma_1$ is in the $135^\circ$ and $45^\circ$ diagonal polarizations, respectively, both of which are linear superpositions of horizontal and vertical polarizations. Consequently, the which-way information of each individual $\gamma_1$ is completely unmarked by the associated $D'_1$ or $D'_2$ outcome, and the photon is said to travel \emph{both} paths simultaneously. Correspondingly, within the confines of either subensemble associated with $D'_1$ or $D'_2$, the detection probabilities of $D_1$ and $D_2$ appear as $\cos^2(\theta/2)$ or $\sin^2(\theta/2)$ in response to the adjustable phase shift $\theta$, manifesting the two-path interference.

Furthermore, If we adjust $\phi'$ to some angle between $0$ and $45^\circ$, the which-way information of each individual $\gamma_1$ is partially (but not completely) marked to a certain degree. Accordingly, within each subensemble associated with $D'_1$ or $D'_2$, the detection probabilities of $D_1$ and $D_2$ appear as partially modulated in response to the adjustable phase shift $\theta$. That is, each subensemble does manifest the two-path interference, but the visibility of the interference pattern is diminished to a certain degree compared to that of the case of $\phi'=45^\circ$. The detection probabilities for the total ensemble remains the same regardless of the value of $\phi'$.

Before the state of $\gamma_2$ is measured by $D'_1$ and $D'_2$, the way how each individual $\gamma_1$ travels the two paths is unknown, or, more precisely, the interpretation of how it travels remains ambiguous. After the measurement of $D'_1$ and $D'_2$, however, this ambiguity is removed, and the way how $\gamma_1$ travels become ascertained.\footnote{\label{foot:concurrence}The fact that the $D'_1$ and $D'_2$ outcome deduces how $\gamma_1$ travels the two paths can be empirically verified by the concurrence counts between $D_1/D_2$ and $D'_1/D'_2$ in the open configuration (i.e., the applied voltage for the EOM is turned off).}
In a sense, the which-way information of each individual $\gamma_1$, which originally can be innocuously presupposed to be either Path 1 or Path 2, can be ``erased'' to a certain degree depending on $\phi'$ by the outcome of the measurement of $D'_1$ and $D'_2$.\footnote{Rigorously speaking, as commented in \ftref{foot:unitary freedom}, the which-way information is neither marked in the first place nor erased in a later time. Rather, it is the ambiguity of interpretations that is removed upon the delayed-choice measurement. It has been argued that no information is erased at all in a quantum eraser and the term ``quantum eraser'' is a misleading misnomer (see e.g., \cite{Qureshi_2020,kastner2019delayed}).}
Note that whether the measurement of $D'_1$ and $D'_2$ is performed before or after the measurement of $D_1$ and $D_2$ is irrelevant. This gives a rather astonishing implication: apparently, the behavior of $\gamma_1$ in the past can be \emph{retroactively} affected by the measurement of $\gamma_2$ performed in the future. What this really means is a matter of philosophy that has sparked intense debate. In this paper, we leave the philosophical question aside but focus on the implementation of a quantum eraser in a quantum circuit, whereby we can investigate the issues of \emph{complementarity} between \emph{visibility} of the interference pattern and \emph{distinguishability} of the which-way information in more depth.

\section{Implementation in a quantum circuit}\label{sec:quantum circuit}
\begin{figure}
\centering
    \includegraphics[width=0.85\textwidth]{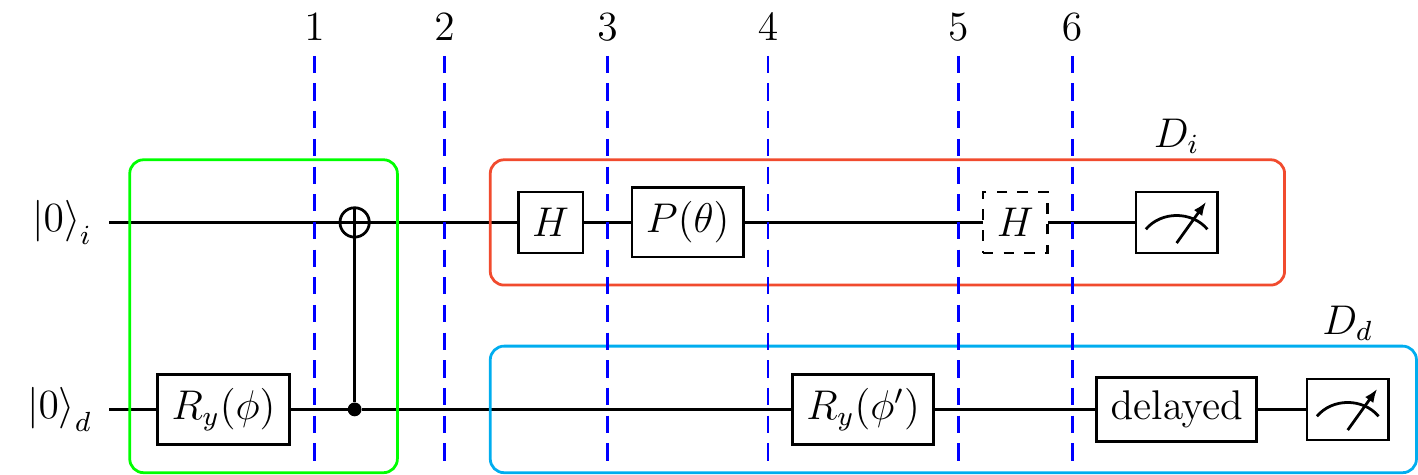}
\caption{Implementation of a delayed-choice quantum eraser in a quantum circuit. The upper right grouped block is analogous to the Mach-Zehnder interferometer in \figref{fig:interferometer}. The lower right grouped block is analogous to the delayed-choice measuring device. The left grouped block is analogous to the BBO module.}
\label{fig:quantum circuit}
\end{figure}

The delayed-choice quantum eraser experiment as illustrated in \figref{fig:interferometer} can be emulated in a quantum circuit as shown in \figref{fig:quantum circuit}, where the phase gate $P(\theta)$ is given by
\begin{equation}
P(\theta)
=
\left(
  \begin{array}{cc}
    1 & 0 \\
    0 & e^{i\theta} \\
  \end{array}
\right),
\end{equation}
and the $R_y(\phi)$ gate is given by
\begin{equation}
R_y(\phi) \equiv e^{-i\phi Y/2}=\cos\frac{\phi}{2}I-i\sin\frac{\phi}{2}Y
=
\left(
  \begin{array}{cc}
    \cos\frac{\phi}{2} & -\sin\frac{\phi}{2} \\
    \sin\frac{\phi}{2} & \cos\frac{\phi}{2} \\
  \end{array}
\right).
\end{equation}

The grouped block in the upper right of \figref{fig:quantum circuit} as a whole emulates the Mach-Zehnder interferometer in the upper part of \figref{fig:interferometer}. If we send a qubit state $\ket{0}$ or $\ket{1}$ into this block, the first Hadamard ($H$) gate transforms it to $1/\sqrt{2}\left(\ket{0}\pm\ket{1}\right)$. After this $H$ gate, the states $\ket{0}$ and $\ket{1}$  can be viewed as analogous to Path 1 and Path 2, respectively. Accordingly, the $H$ gate is analogous to the first PBS in \figref{fig:interferometer}, which splits the incoming photon into two paths. The $P(\theta)$ gate is analogous to the adjustable phase-shift plate $\theta$, which adds a relative phase $e^{i\theta}$ to Path 2 (i.e., $\ket{1}$ by analogy). The second (dashed) $H$ gate is analogous to the module composed of the second PBS, the half-wave plate, and the EOM, in \figref{fig:interferometer}, which recombines the states of Path 1 (i.e., $\ket{0}$) and Path 2 (i.e., $\ket{1}$). Finally, the meter $D_i$ is analogous to the Wollaston prism together with the detectors $D_1$ and $D_2$.\footnote{We label the objects associated in the upper quantum wire with the superscript or subscript ``i'' for ``interference'' and those in the lower wire with ``d'' for ``delayed-choice''.} The $0/1$ readouts of $D_i$ are analogous to the signals registered in $D_1$ and $D_2$, respectively.\footnote{\label{foot:diagonal polarization}Prior to the first $H$ gate, the qubit states $\ket{0}_i$ and $\ket{1}_i$ are analogous to the photon states of $45^\circ$ and $135^\circ$ diagonal polarizations, which become $1/\sqrt{2}\left(\ket{\mathrm{Path\ 1}}\pm\ket{\mathrm{Path\ 2}}\right)$ after entering the first PBS. After the second $H$ gate, the qubit states $\ket{0}_i$ and $\ket{1}_i$ are analogous to the photon states of horizontal and vertical polarizations, which enter $D_1$ and $D_2$, respectively. In the Mach-Zehnder interferometer as shown in \figref{fig:interferometer}, an entering beam is split into two beams, which are recombined and strike either $D_1$ or $D_2$ in the end. By contrast, in the quantum circuit analogy, there is only one qubit throughout the whole ``interferometer'', which is measured with the $0/1$ readouts by $D_i$ in the end.} Furthermore, if the dashed $H$ gate is removed, the whole block then emulates the open configuration of the Mach-Zehnder interferometer.

On the other hand, the grouped block in the lower right of \figref{fig:quantum circuit} as a whole emulates the delayed-choice measuring device in the lower part of \figref{fig:interferometer}. Tuning $\phi'$ for the $R_y(\phi')$ gate is analogous to adjusting the orientation angle $\phi'$ for the Wollaston prism WP($\phi'$).\footnote{However, the value of $\phi'$ in $R_y(\phi')$ is not to be directly identified with the value of $\phi'$ in WP($\phi'$) by the analogy. In accordance with the comment in \ftref{foot:diagonal polarization}, the qubit states $\ket{0}$ and $\ket{1}$ before the $R_y(\phi')$ gate are analogous to the photon states of $45^\circ$ and $135^\circ$ diagonal polarizations. Consequently, in particular, $R_y(\phi'=0)$ corresponds to WP($\phi'=\pi/4$), and $R_y(\phi'=\pi/2)$ corresponds to WP($\phi'=0$).} The $0/1$ readouts of the meter $D_d$ are analogous to the signals registered in $D'_1$ and $D'_2$, respectively. We also inserted a delayed gate to ensure that the measurement by $D_d$ is performed after the measurement by $D_i$.\footnote{Note that the upper right and lower right grouped blocks do not touch each other at all. Therefore, the exact position of the $R_y(\phi')$ gate relative to the gates of the upper grouped block is not important. In fact, the $R_y(\phi')$ gate can be positioned even after the delayed gate. It is only convenient for illustration and calculation that the $R_y(\phi')$ gate is depicted in this particular position.}

Meanwhile, the grouped block in the left of \figref{fig:quantum circuit} as a whole emulates the BBO module in \figref{fig:interferometer}, which provides a pair of entangled photons. Feeding $\ket{0}$ to the two quantum wires, the grouped block provides the state $\ket{\psi_2}$ given by \eqref{psi2}. If we set $\phi=\pi/2$, $\ket{\psi_2}=1/\sqrt{2}\left(\ket{00}+\ket{11}\right)$ gives a pair of fully entangled qubits, analogous to the entangled photons produced the BBO module.
Furthermore, we can adjust $\phi$ to any value and thus provide a pair of qubits with any arbitrary degree of entanglement. This adjustment enables us to investigate consequences of the delayed-choice eraser in a broader setting with adjustable entanglement between the two paired quantons, which cannot be easily implemented in an optical experiment.

In the quantum circuit implemented in \figref{fig:quantum circuit}, the quantum state at each slice can be straightforwardly calculated as
\begin{subequations}
\begin{eqnarray}
\ket{\psi_1} &=& \ket{0}_i\left(\cos\frac{\phi}{2}\ket{0}_d + \sin\frac{\phi}{2}\ket{1}_d\right), \\
\label{psi2}
\ket{\psi_2} &=& \cos\frac{\phi}{2}\ket{0}_i\ket{0}_d + \sin\frac{\phi}{2}\ket{1}_i\ket{1}_d, \\
\ket{\psi_3} &=& \frac{\cos(\phi/2)}{\sqrt{2}}(\ket{0}_i+\ket{1}_i)\ket{0}_d
                +\frac{\sin(\phi/2)}{\sqrt{2}}(\ket{0}_i-\ket{1}_i)\ket{1}_d, \\
\label{psi4}
\ket{\psi_4} &=& \frac{\cos(\phi/2)}{\sqrt{2}}(\ket{0}_i+e^{i\theta}\ket{1}_i)\ket{0}_d
                +\frac{\sin(\phi/2)}{\sqrt{2}}(\ket{0}_i-e^{i\theta}\ket{1}_i)\ket{1}_d, \\
\label{psi5}
\ket{\psi_5} &=&
\frac{1}{\sqrt{2}}\left(\cos\frac{\phi+\phi'}{2}\ket{0}_i+e^{i\theta}\cos\frac{\phi-\phi'}{2}\ket{1}_i\right)
\ket{0}_d \nonumber\\
&& \quad \mbox{} + \frac{1}{\sqrt{2}}\left(\sin\frac{\phi+\phi'}{2}\ket{0}_i-e^{i\theta}\sin\frac{\phi-\phi'}{2}\ket{1}_i\right)
\ket{1}_d, \\
\label{psi6}
\ket{\psi_6} &=&
\frac{1}{2}\left(
\left[\cos\frac{\phi+\phi'}{2}+e^{i\theta}\cos\frac{\phi-\phi'}{2}\right]\ket{0}_i +
\left[\cos\frac{\phi+\phi'}{2}-e^{i\theta}\cos\frac{\phi-\phi'}{2}\right]\ket{1}_i\right)
\ket{0}_d \nonumber\\
&& \mbox{} +
\frac{1}{2}\left(
\left[\sin\frac{\phi+\phi'}{2}-e^{i\theta}\sin\frac{\phi-\phi'}{2}\right]\ket{0}_i +
\left[\sin\frac{\phi+\phi'}{2}+e^{i\theta}\sin\frac{\phi-\phi'}{2}\right]\ket{1}_i\right)
\ket{1}_d. \qquad
\end{eqnarray}
\end{subequations}

If we focus on the qubit $\ket{q}_i$ of the upper (``i'') quantum wire, it is described by the reduced density matrix $\rho^{(i)}_n:= \Tr_{\ket{q}_d}(\ket{\psi_n}\bra{\psi_n})$ traced out over the qubit $\ket{q}_d$ of the lower (``d'') quantum wire. The reduced density matrix $\rho^{(i)}_n$ at each slice is given by
\begin{subequations}
\begin{eqnarray}
\rho^{(i)}_2
&=&
\left(
  \begin{array}{cc}
     \cos^2\frac{\phi}{2} & 0 \\
     0 & \sin^2\frac{\phi}{2}
  \end{array}
\right), \\
\rho^{(i)}_3
&=&
\frac{1}{2}
\left(
  \begin{array}{cc}
     1 & \cos\phi \\
     \cos\phi & 1
  \end{array}
\right), \\
\rho^{(i)}_4 \equiv \rho^{(i)}_5 = P(\theta)\rho^{(i)}_3P(\theta)^\dag
&=&
\frac{1}{2}
\left(
  \begin{array}{cc}
     1 & \cos\phi\, e^{-i\theta} \\
     \cos\phi\, e^{i\theta} & 1
  \end{array}
\right), \\
\rho^{(i)}_6
= H \rho^{(i)}_5 H
&=&
\frac{1}{2}
\left(
  \begin{array}{cc}
     1+\cos\phi\cos\theta & i\cos\phi\sin\theta \\
     -i\cos\phi\sin\theta & 1-\cos\phi\cos\theta
  \end{array}
\right).
\end{eqnarray}
\end{subequations}
By adjusting $\phi$, the reduced density matrix can provide any degree of entanglement, quantified by the \emph{entropy of entanglement}
\begin{equation}
S(\rho^{(i)}) \equiv -\Tr(\rho^{(i)}\log\rho^{(i)})
= -\cos^2\frac{\phi}{2}\log\left(\cos^2\frac{\phi}{2}\right)
-\sin^2\frac{\phi}{2}\log\left(\sin^2\frac{\phi}{2}\right).
\end{equation}
Alternatively, $\rho^{(i)}$ is said to have the \emph{degree of purity}
\begin{equation}\label{purity}
\Tr\,(\rho^{(i)})^2 = \frac{1+\cos^2\phi}{2}
\end{equation}
for any arbitrary value. The entropy of entanglement increases as the degree of purity decreases.

On the other hand, if we focus on the ``d'' qubit $\ket{q}_d$, it is described by the reduced density matrix $\rho^{(d)}_n:= \Tr_{\ket{q}_i}(\ket{\psi_n}\bra{\psi_n})$ traced out over the ``i'' qubit. The reduced density matrix $\rho^{(d)}_n$ at each slice is given by
\begin{subequations}
\begin{eqnarray}
\rho^{(d)}_2\equiv\rho^{(d)}_3\equiv\rho^{(d)}_4
&=&
\left(
  \begin{array}{cc}
     \cos^2\frac{\phi}{2} & 0 \\
     0 & \sin^2\frac{\phi}{2}
  \end{array}
\right), \\
\label{rho5 d}
\rho^{(d)}_5 = R_y(\phi')\rho^{(d)}_4R_y(\phi')^\dag
&=&
\frac{1}{2}
\left(
  \begin{array}{cc}
     \cos^2\frac{\phi+\phi'}{2}+ \cos^2\frac{\phi-\phi'}{2} & \cos\phi\sin\phi' \\
     \cos\phi\sin\phi' & \sin^2\frac{\phi+\phi'}{2}+ \sin^2\frac{\phi-\phi'}{2}
  \end{array}
\right) \nonumber\\
&=&
\frac{1}{2}
\left(
  \begin{array}{cc}
     1+\cos\phi\cos\phi' & \cos\phi\sin\phi' \\
     \cos\phi\sin\phi' & 1-\cos\phi\cos\phi'
  \end{array}
\right).
\end{eqnarray}
\end{subequations}
Note that the entropy of entanglement $S(\rho^{(d)})$ and the degree of purity $\Tr\,(\rho^{(d)})^2$ are the same as those of $\rho^{(i)}$.\footnote{If a composite $AB$ system is in a pure state, the Schmidt decomposition implies that the density matrices $\rho^A$ and $\rho^B$ for the $A$ and $B$ subsystems, respectively, have the same eigenvalues.}

\section{Visibility and distinguishability}\label{sec:V and D}
The states and density matrices at each slice in \figref{fig:quantum circuit} have been calculated in the previous section, we can now study visibility of the interference pattern measured by $D_i$, and distinguishability of the which-way information inferred from the outcome measured by $D_d$. We will investigate complementarity relations between visibility and distinguishability from three different perspectives: (i) in view of the total ensemble of all events, (ii) in view of the subensembles respectively associated with the readouts 0 and 1 in $D_d$, and (iii) in view of the average results averaged over the two subensembles.

\subsection{Total ensemble perspective}
By repeating the experiment many times (for a given phase shift $\theta$), we obtain an ensemble of accumulated counts of individual readouts.
Among the ensemble, the probability $p(0_i)$ of having the readout 0 in $D_i$ in \figref{fig:quantum circuit} and the probability $p(1_i)$ of having the readout 1 are respectively given by
\begin{subequations}\label{p i total}
\begin{eqnarray}
  p(0_i) &=& \Tr\left(\ket{0_i}\bra{0_i}\rho^{(i)}_6\right) = \frac{1}{2}\left(1+\cos\phi\cos\theta\right), \\
  p(1_i) &=& \Tr\left(\ket{1_i}\bra{1_i}\rho^{(i)}_6\right) = \frac{1}{2}\left(1-\cos\phi\cos\theta\right).
\end{eqnarray}
\end{subequations}
Similarly, the probability $p(0_d)$ of having the readout 0 in $D_d$ and the probability $p(1_d)$ of having the readout 1 are respectively given by
\begin{subequations}\label{p d total}
\begin{eqnarray}
  p(0_d) &=& \Tr\left(\ket{0_d}\bra{0_d}\rho^{(d)}_5\right) = \frac{1}{2}\left(1+\cos\phi\cos\phi'\right), \\
  p(1_d) &=& \Tr\left(\ket{1_d}\bra{1_d}\rho^{(d)}_5\right) = \frac{1}{2}\left(1-\cos\phi\cos\phi'\right).
\end{eqnarray}
\end{subequations}

Generally, the detection probabilities in \eqref{p i total} manifest the two-path interference as modulated in response to the phase shift $\theta$. However, the interference pattern is less significant when $|\cos\phi|$ becomes smaller. To quantify the visibility of the interference pattern, we first define the contrast of the interference pattern as
\begin{equation}\label{eq:totcontrast}
\mathcal{C} := p(0_i) - p(1_i) \equiv 2p(0_i)-1
= \cos\phi\cos\theta.
\end{equation}
The \emph{visibility} of the interference is then defined as
\begin{equation}\label{V}
\mathcal{V} := \frac{\max_\theta p(0_i)-\min_\theta p(0_i)}
{\max_\theta p(0_i)+\min_\theta p(0_i)}
\equiv
\frac{\max_\theta\mathcal{C}-\min_\theta\mathcal{C}}
{2+\max_\theta\mathcal{C}+\min_\theta\mathcal{C}}
= |\cos\phi|.
\end{equation}
This tells that the visibility of interference increases as the purity of $\rho^{(i)}$ increases.

Accordingly to the \emph{complementarity principle}, the more significant the interference pattern is, the less certain the which-way information can be inferred. To quantify how much the which-way information is deducible, we first compute the probability of correctly identifying which-way information based on the readout measured by $D_d$. The which-way information of the ``i'' qubit can be explicitly measured in the open configuration where the dashed $H$ gate in \figref{fig:quantum circuit} is removed. In the open configuration, the two ``paths'' $\ket{0}$ and $\ket{1}$ register the readouts 0 and 1 separately in $D_i$, so the readout of $D_i$ completely tells the which-way information.
In the closed configuration, on the other hand, the which-way information cannot be determined from the readout of $D_i$, but it can be indirectly inferred with a certain degree of certainty from the readout of $D_d$. Let us adopt an identifying strategy as follows: the which-way information of $\ket{q}_i$ is guessed to be $\ket{0}$ if $D_d$ yields 0, and $\ket{1}$ if $D_d$ yields 1. The probability of successfully guessing the which-way information can be empirically computed from the concurrence counts between the readouts of $D_i$ and $D_d$ in the open configuration (recall \ftref{foot:concurrence}).
Mathematically, the probability of success is computed as
\begin{eqnarray}\label{p succ}
p_\mathrm{succ} &=& p(0_d)\,p(0_i|0_d) + p(1_d)\,p(1_i|1_d)
\equiv p(0_i,0_d) + p(1_i,1_d)\nonumber\\
&=&
\big|\langle(\ket{0}_i\ket{0}_d)|\psi_5\rangle\big|^2
+
\big|\langle(\ket{1}_i\ket{1}_d)|\psi_5\rangle\big|^2
\nonumber\\
&=& \frac{1}{2}\left(\cos^2\frac{\phi+\phi'}{2}
+\sin^2\frac{\phi-\phi'}{2}\right)
=\frac{1}{2}-\frac{1}{2}\sin\phi\sin\phi'.
\end{eqnarray}
If $p_\mathrm{succ}=1/2$, the strategy just provides a random guess. If $1/2<p_\mathrm{succ}\leq1$, the which-way information can be inferred with a certain certainty. If $0\leq p_\mathrm{succ}<1/2$, it just means that we should have adopted the strategy the other way around (i.e., $\ket{q}_i$ is identified to be $\ket{1}$ if $D_d$ yields 0, and $\ket{0}$ if $D_d$ yields 1).
Correspondingly, the \emph{distinguishability} by the strategy is defined as $2\,p_\mathrm{succ}-1$ and given by
\begin{equation}\label{D}
\mathcal{D}:= 2\,p_\mathrm{succ}-1
=-\sin\phi\sin\phi'.
\end{equation}
Again, if $\mathcal{D}<0$, it means that the strategy should have been the other way around.
In the closed configuration, the certainty of deducing the which-way information of $\ket{q}_i$ from the readout of $D_d$ is said to be described by the distinguishability $\mathcal{D}$.\footnote{\label{foot:special cases}In the special case that $\cos\phi\cos\phi'=1$, we have $p(0_d)=1$ and $p(1_d)=0$, and thus $p(0_i|1_d)$ and $p(1_i|1_d)$ appearing in \eqref{p succ} are ill defined. Similarly, in the special case that $\cos\phi\cos\phi'=-1$, we have $p(0_d)=0$ and $p(1_d)=1$, and thus $p(0_i|0_d)$ and $p(1_i|0_d)$ are ill defined. Nevertheless, for both special cases, $p_\mathrm{succ}$ in \eqref{p succ} and thus $\mathcal{D}$ in \eqref{D} are still well defined, as $p_\mathrm{succ}$ simply reduces to $p(1_d)p(1_i|1_d)$ or $p(0_d)p(0_i|0_d)$, respectively, when $p(0_d)=0$ or $p(1_d)=0$.}

From \eqref{V} and \eqref{D}, we obtain the complementarity relation
\begin{equation}\label{V D duality}
\mathcal{V}^2+\mathcal{D}^2 = \cos^2\phi + \sin^2\phi\sin^2\phi' \leq 1,
\end{equation}
which is in agreement with \eqref{wave-particle duality}.
This inequality can be saturated by choosing $\phi'=\pm\pi/2$, at which $|\mathcal{D}|$ gives the optimal value $|\sin\phi|$ for a given $\phi$. We will discuss \eqref{V D duality} in more depth in \secref{sec:comparison}.

\subsection{Subensemble perspective}\label{sec:subensemble}
On the other hand, instead of the total ensemble, we can consider the subensemble of the events associated with the readout 0 in $D_d$ and the subensemble associated with the readout 1 in $D_d$ \emph{separately}. Within either of the two subensembles (labeled with ``$0_d$'' and ``$1_d$'' respectively), the which-way information of $\ket{q}_i$ can be partially or completely erased. Accordingly, the interference pattern of $\ket{q}_i$ within the confines of either subensemble could exhibit higher visibility than that of the total ensemble.

According to \eqref{psi6}, for the events corresponding to $0_d$, the wavefunction of $\ket{q}_i$ is collapsed into
\begin{equation}
\ket{\psi}_i \propto \left(\cos\frac{\phi+\phi'}{2}+e^{i\theta}\cos\frac{\phi-\phi'}{2}\right)\ket{0}_i +
\left(\cos\frac{\phi+\phi'}{2}-e^{i\theta}\cos\frac{\phi-\phi'}{2}\right)\ket{1}_i.
\end{equation}
Within the $0_d$ subensemble, the probability of having the readout 0 in $D_i$ and the probability of having the readout 1 are given respectively by
\begin{subequations}\label{p 0d}
\begin{eqnarray}
p(0_i|0_d) &=& \frac{\big|\langle0\ket{\psi}\big|^2}{\big|\langle\psi\ket{\psi}\big|^2}
= \frac{1+\cos\phi\cos\phi'+(\cos\phi+\cos\phi')\cos\theta}
{2\left(1+\cos\phi\cos\phi'\right)}, \\
p(1_i|0_d) &=& \frac{\big|\langle1\ket{\psi}\big|^2}{\big|\langle\psi\ket{\psi}\big|^2}
= \frac{1+\cos\phi\cos\phi'-(\cos\phi+\cos\phi')\cos\theta}
{2\left(1+\cos\phi\cos\phi'\right)}.
\end{eqnarray}
\end{subequations}
Compared with \eqref{p i total} for the interference pattern of the total ensemble, where the modulation in response to $\theta$ vanishes if $\cos\phi=0$, the modulation can be restored in \eqref{p 0d} by tuning $\phi'$.
The contrast of the interference pattern for the $0_d$ subensemble is defined and given by
\begin{equation}\label{C 0d}
\mathcal{C}_{0_d} := p(0_i|0_d) - p(1_i|0_d)
= \frac{\cos\phi+\cos\phi'}{1+\cos\phi\cos\phi'}
\cos\theta.
\end{equation}
The visibility of the interference corresponding to $0_d$ is then defined and given by
\begin{equation}\label{V 0d}
\mathcal{V}_{0_d} :=
\frac{\max_\theta p(0_i|0_d)-\min_\theta p(0_i|0_d)}
{\max_\theta p(0_i|0_d)+\min_\theta p(0_i|0_d)}
\equiv
\frac{\max_\theta\mathcal{C}_{0_d}-\min_\theta\mathcal{C}_{0_d}}
{2+\max_\theta\mathcal{C}_{0_d}+\min_\theta\mathcal{C}_{0_d}}
=
\frac{|\cos\phi+\cos\phi'|}{1+\cos\phi\cos\phi'}.
\end{equation}

Similarly, for the events corresponding to $1_d$, the wavefunction of $\ket{q}_i$ is collapsed into
\begin{equation}
\ket{\psi}_i \propto
\left(\sin\frac{\phi+\phi'}{2}+e^{i\theta}\sin\frac{\phi-\phi'}{2}\right)\ket{0}_i +
\left(\sin\frac{\phi+\phi'}{2}-e^{i\theta}\sin\frac{\phi-\phi'}{2}\right)\ket{1}_i.
\end{equation}
Within the $1_d$ subensemble, the probability of having the readout 0 in $D_i$ and the probability of having the readout 1 are given respectively by
\begin{subequations}\label{p 1d}
\begin{eqnarray}
p(0_i|1_d) &=& \frac{\big|\langle0\ket{\psi}\big|^2}{\big|\langle\psi\ket{\psi}\big|^2}
= \frac{1-\cos\phi\cos\phi'+(\cos\phi'-\cos\phi)\cos\theta}
{2\left(1-\cos\phi\cos\phi'\right)}, \\
p(1_i|1_d) &=& \frac{\big|\langle1\ket{\psi}\big|^2}{\big|\langle\psi\ket{\psi}\big|^2}
= \frac{1-\cos\phi\cos\phi'-(\cos\phi'-\cos\phi)\cos\theta}
{2\left(1-\cos\phi\cos\phi'\right)}.
\end{eqnarray}
\end{subequations}
The contrast of the interference pattern for the $1_d$ subensemble is defined and given by
\begin{equation}
\mathcal{C}_{1_d} = p(0_i|1_d) - p(1_i|1_d)
= \frac{\cos\phi'-\cos\phi}{1-\cos\phi\cos\phi'}
\cos\theta.
\end{equation}
The visibility of the interference corresponding to $0_d$ is then defined and given by
\begin{equation}\label{V 1d}
\mathcal{V}_{1_d} :=
\frac{\max_\theta p(0_i|1_d)-\min_\theta p(0_i|1_d)}
{\max_\theta p(0_i|1_d)+\min_\theta p(0_i|1_d)}
\equiv
\frac{\max_\theta\mathcal{C}_{1_d}-\min_\theta\mathcal{C}_{1_d}}
{2+\max_\theta\mathcal{C}_{1_d}+\min_\theta\mathcal{C}_{1_d}}
=\frac{|\cos\phi'-\cos\phi|}{1-\cos\phi\cos\phi'}.
\end{equation}

We can also consider and define the \emph{distinguishabilities} $\mathcal{D}_{0_d}$ and $\mathcal{D}_{1_d}$ \emph{within} the $0_d$ subensemble and the $1_d$ subensemble, respectively.
Computing $p(0_i|0_d)$ and $p(1_i|a_d)$ from $\ket{\psi_5}$ given in \eqref{psi5}, we have
\begin{subequations}\label{D 0d 1d}
\begin{eqnarray}
\mathcal{D}_{0_d} &:=& 2\,p_\mathrm{succ}^{(0_d)}-1 = 2\,p(0_i|0_d)-1
= \frac{-\sin\phi\sin\phi'}{1+\cos\phi\cos\phi'}, \\
\mathcal{D}_{1_d} &:=& 2\,p_\mathrm{succ}^{(1_d)}-1 = 2\,p(1_i|1_d)-1
= \frac{\sin\phi\sin\phi'}{1-\cos\phi\cos\phi'},
\end{eqnarray}
\end{subequations}
where $p_\mathrm{succ}^{(0_d)}$ and $p_\mathrm{succ}^{(1_d)}$ are the probabilities of successfully identifying
the which-way information of $\ket{q}_i$ \emph{within} the $0_d$ subensemble and the $1_d$ subensemble, respectively.\footnote{\label{foot:special cases 2}As remarked in \ftref{foot:special cases}, we have to pay special attention to the two special cases. In the case that $\cos\phi\cos\phi'=1$, there are no events in the $1_d$ subensemble (i.e., $p(1_d)=0$), and $\mathcal{D}_{1_d}$ and $\mathcal{V}_{1_d}$ are both ill defined. Similarly, in the case that $\cos\phi\cos\phi'=-1$, there are no events in the $0_d$ subensemble (i.e., $p(0_d)=0$), and $\mathcal{D}_{0_d}$ and $\mathcal{V}_{0_d}$ are both ill defined.}

Given \eqref{V 0d}, \eqref{V 1d}, and \eqref{D 0d 1d}, it is easy to show that
\begin{subequations}\label{V D duality sub}
\begin{eqnarray}
\mathcal{V}_{0_d}^2+\mathcal{D}_{0_d}^2 &=& 1, \\
\mathcal{V}_{1_d}^2+\mathcal{D}_{1_d}^2 &=& 1.
\end{eqnarray}
\end{subequations}
It is rather curious that, \emph{within} either of the two subensemble, the complementarity relation \eqref{wave-particle duality} is saturated. This aspect will be further discussed in \secref{sec:comparison}.

For a typical delayed-choice quantum eraser with full entanglement (such as the optical experiment in \figref{fig:interferometer}), we have $\phi=\pm\pi/2$ or equivalently $\cos\phi=0$. Correspondingly, $\mathcal{V}=0$ in \eqref{V}, and both $\mathcal{V}_{d_0}$ and $\mathcal{V}_{d_1}$ are greater than $\mathcal{V}$ unless $\cos\phi'=0$. The visibility is said to be ``recovered'' to a certain degree as long as $|\mathcal{D}_{d_0}|$ and $|\mathcal{D}_{d_1}|$ remain less than 1 (i.e., $\cos\phi'\neq0$).
In a generic case of partial entanglement (i.e., $\phi\neq\pm\pi/2$ or equivalently $\cos\phi\neq0$), it is possible that one of $\mathcal{V}_{d_0}$ and $\mathcal{V}_{d_1}$ becomes smaller than $\mathcal{V}$. Nevertheless, on average, the average visibility is always greater than or equal to the visibility $\mathcal{V}$ for the total ensemble as will be shown in the next subsection.
In the case of $\phi'=0$, the visibility is restored to unity, i.e., $\mathcal{V}_{0_d}=\mathcal{V}_{1_d}=1$, while the distinguishability completely vanishes, i.e., $\mathcal{D}_{0_d}=\mathcal{D}_{1_d}=0$, regardless of $\phi$.

\subsection{Average perspective}
With reference to the two subensembles, we can define and compute the \emph{average visibility} as
\begin{eqnarray}\label{V avg}
\mathcal{V}_\mathrm{avg} &:=& p(0_d)\mathcal{V}_{0_d} + p(1_d)\mathcal{V}_{1_d} \nonumber\\
&=&
\frac{1}{2}|\cos\phi+\cos\phi'|+\frac{1}{2}|\cos\phi-\cos\phi'| \nonumber\\
&=&
\max\left(|\cos\phi|, |\cos\phi'|\right),
\end{eqnarray}
where \eqref{p d total}, \eqref{V 0d}, and \eqref{V 1d} have been used.
Similarly, we can define and compute the \emph{average distinguishability} as
\begin{eqnarray}\label{D avg}
\mathcal{D}_\mathrm{avg} &:=& p(0_d)\mathcal{D}_{0_d} + p(1_d)\mathcal{D}_{1_d} \nonumber\\
&=& 2p(0_d)p(0_i|0_d)-p(0_d) +2p(1_d)p(1_i|1_d)-p(1_d) \nonumber\\
&=& 2p_\mathrm{succ} -1 \equiv \mathcal{D} = -\sin\phi\sin\phi',
\end{eqnarray}
where \eqref{p succ} and \eqref{D} have been used.\footnote{\label{foot:special cases 3}As remarked in \ftref{foot:special cases 2}, in the special case that $\cos\phi\cos\phi'=1$, $\mathcal{D}_{1_d}$ and $\mathcal{V}_{1_d}$ are ill defined; in the special case that $\cos\phi\cos\phi'=-1$, $\mathcal{D}_{0_d}$ and $\mathcal{V}_{0_d}$ are ill defined. Nevertheless, for both special cases, $\mathcal{V}_\mathrm{avg}$ and $\mathcal{D}_\mathrm{avg}$ are still well defined, because the ill defined part is multiplied by $p(1_d)=0$ or $p(0_d)=0$ in \eqref{V avg} and \eqref{D avg}.}
Note that the average distinguishability $\mathcal{D}_\mathrm{avg}$ is identical to $\mathcal{D}$ as expected, since the probability of successfully identifying the which-way information is independent of how the total ensemble is divided into subgroups.
From the average point of view, we again have the complementarity relation,
\begin{equation}\label{V D duality avg}
\mathcal{V}_\mathrm{avg}^2+\mathcal{D}_\mathrm{avg}^2 = \max\left(\cos^2\phi,\cos^2\phi'\right) + \sin^2\phi\sin^2\phi' \leq 1.
\end{equation}
Note that, whereas $\mathcal{D}_\mathrm{avg}$ is identical to $\mathcal{D}$, the average visibility $\mathcal{V}_\mathrm{avg}$ given by \eqref{V avg} is equal to or greater than the original $\mathcal{V}$ given by \eqref{V}, depending on $\phi'$. Nevertheless, the complementarity relation \eqref{wave-particle duality} still holds for $\mathcal{V}_\mathrm{avg}$ and $\mathcal{D}_\mathrm{avg}$. This will be discussed further in \secref{sec:comparison}.

\section{Comparison to established perspectives in the literature}\label{sec:comparison}
In this section, we will study and discuss the complementarity relations obtained in the previous section from established perspectives in the literature, particularly in view of the entropic framework proposed in \cite{PhysRevA.93.062111} and in view of the triality relation proposed by \cite{Jakob2010}.

\subsection{Entropic framework}
The entropic framework of \cite{PhysRevA.93.062111} provides a generic scheme for formulating wave-particle duality relations in arbitrary multipath interferometers from the uncertainty relations for the min- and max-entropies. The framework considers a generic $n$-path interferometer for single quantum particles. After the $n$ paths are recombined, the particle is detected at one of the detectors, giving rise to an interference pattern for the accumulated count of individual signals. On the other hand, the particle also interacts with some environment, upon which the imprint can be used to infer the which-way information of the particle. The schematic setting is depicted in Fig.~1 of \cite{PhysRevA.93.062111}.
The which-way distinguishability $\mathcal{D}$ is defined in Eq.~(16) in \cite{PhysRevA.93.062111} as
\begin{equation}\label{D n-path}
\mathcal{D} := \frac{n\,p_\mathrm{guess}(Z|E)-1}{n-1},
\end{equation}
where $Z$ is the random variable for which-way information, and $p_\mathrm{gusss}(Z|E)$ is the probability of guessing $Z$ correctly given the outcome of the \emph{optimal} measurement on the environment $E$.
On the other hand, for the case that the interferometer is symmetric (i.e., a particle traveling along a well-defined path inside the interferometer arrives at each of the detectors with an equal probability), the interference visibility $\mathcal{V}$ is defined in Eq.~(20) in \cite{PhysRevA.93.062111} as
\begin{equation}\label{V n-path}
\mathcal{V} := \frac{n\,p_\mathrm{guess}^{\max}(C)-1}{n-1},
\end{equation}
where $p_\mathrm{guess}^{\max}(C):= \max_{\vec{\phi}}p_\mathrm{guess}(C)$ is the maximum of $p_\mathrm{guess}(C)$ over all configurations of phase shifts $\vec{\phi}=(\phi_1,\phi_2,\dots,\phi_n)$ applied to the $n$ paths, and $p_\mathrm{guess}(C)$ is the probability of correctly guessing which detector clicks.
The uncertainty relation for the min- and max-entropies then leads to the generalized wave-particle duality relation
\begin{equation}\label{V D duality entropic view}
\mathcal{V}^2+\mathcal{D}^2 \leq 1
\end{equation}
as given in Eq.~(21) in \cite{PhysRevA.93.062111}.

For a two-path interferometer, \eqref{D n-path} reduces to the form
\begin{equation}\label{D 2-path}
\mathcal{D} = 2\, p_\mathrm{gusss}(Z|E)-1,
\end{equation}
and \eqref{V n-path} reduces to the form
\begin{equation}\label{V 2-path}
\mathcal{V} = \frac{p_{C=1}^{\max}-p_{C=1}^{\min}}{p_{C=1}^{\max}+p_{C=1}^{\min}},
\end{equation}
where $p_{C=1}^{\max}:=\max_{\vec{\phi}}(p_{C=1})$, $p_{C=1}^{\min}:=\min_{\vec{\phi}}(p_{C=1})$, and $p_{C=1}$ is the probability that the detector $D_1$ clicks. Note that \eqref{V 2-path} is equivalent to the definition adopted in \eqref{V} for the quantum circuit implementation. However, \eqref{D 2-path} is slightly different from the definition adopted in \eqref{D}, as the former is related to the probability of correctly guessing the which-way information given the outcome of the \emph{optimal} measurement on the environment, while the latter is related to the probability given the outcome of the $D_d$ measurement with a specific value of $\phi'$. As the optimal $\mathcal{D}$ is of course greater than a nonoptimal $\mathcal{D}$, this explains why we have $\mathcal{V}^2+\mathcal{D}^2 \leq 1$ in \eqref{V D duality}.
Also note that if we choose $\phi'=\pm\pi/2$, the distinguishability $\mathcal{D}$ in \eqref{D} becomes optimal, and the complementarity relation \eqref{V D duality} is saturated (we will discuss this further in the next subsection).

Furthermore, the entropic framework of \cite{PhysRevA.93.062111} also gives a wave-particle duality from the viewpoint of quantum erasure. Consider a POVM measurement $\mathbb{Y}=\{\mathbb{Y}_y\}$ performed on the environment. This discriminates all measurement events of the system into subensembles associated with the different outcomes $y$. For the subensemble associated with $y$, one can define the distinguishability and visibility as before by treating the subensemble as the total ensemble and denote them as $\mathcal{D}(\mathbb{Y}_y)$ and $\mathcal{V}(\mathbb{Y}_y)$, respectively. Correspondingly, one can then define the \emph{average} distinguishability and visibility as
\begin{subequations}\label{D and V erasure}
\begin{eqnarray}
  \mathcal{D}(\mathbb{Y}) &:=& \sum_y p_y\mathcal{D}(\mathbb{Y}_y), \\
  \mathcal{V}(\mathbb{Y}) &:=& \sum_y p_y\mathcal{V}(\mathbb{Y}_y),
\end{eqnarray}
\end{subequations}
as given in Eq.~(27) in \cite{PhysRevA.93.062111}. For any choice of $\mathbb{Y}$, the uncertainty relation for the min- and max-entropies again implies
\begin{equation}\label{}\label{V D duality erasure}
  \mathcal{V}(\mathbb{Y})^2 + \mathcal{D}(\mathbb{Y})^2 \le 1
\end{equation}
as given in Eq.~(28) in \cite{PhysRevA.93.062111}.
Note that the average distinguishability and visibility defined in \eqref{D and V erasure} is equivalent to \eqref{V avg} and \eqref{D avg}. Therefore, the complementarity relation $\mathcal{V}_\mathrm{avg}^2+\mathcal{D}_\mathrm{avg}^2\le1$ obtained in \eqref{V D duality avg} is just a special case of \eqref{V D duality erasure}.

It should be remarked that, there are two very different scenarios for deducing which-way information from the environment. The first is to bring the interferometer into contact with the environment. The second, by contrast, does not invoke any direct contact with the environment at all, but instead the ``interaction'' between the interferometer and the environment is through entanglement.
The work of \cite{PhysRevA.104.032223} provides the first scenario implemented in a quantum circuit (see Fig.~1 thereof).
In our work, we consider the same quantum circuit in analogy to a two-path Mach-Zehnder interferometer, but the which-way information is inferred by the second scenario. The results of both works conform with the wave-particle complementarity relations given by \cite{PhysRevA.93.062111}, concretely showing that the entropic framework of \cite{PhysRevA.93.062111} is applicable to generic settings regardless of whether the system is in direct contact with the environment or is entangled with the environment.

However, as our implementation has some peculiar features due to entanglement that are not addressed in the framework of \cite{PhysRevA.104.032223}. We will discuss them in the next subsection.

\subsection{Triality relation}\label{sec:triality relation}
Within either of the subensembles associated with $0_d$ and $1_d$, the wave-particle complementarity relation is also satisfied as shown in \eqref{V D duality sub}, but it is curious why the complementarity relation is saturated. This can be understood by a simple calculation as provided in \appref{app:comp}, but it is more instructive to understand this feature as a natural consequence from the triality relation for bipartite systems proposed in \cite{Jakob2010}.

A general bipartite state of two-dimensional (i.e., qubit) systems is given by
\begin{equation}
\ket{\Theta} = a \ket{00}+b\ket{01}+c\ket{10}+d\ket{11},
\end{equation}
with $|a|^2+|b|^2+|c|^2+|d|^2=1$.
The bipartite state exhibits one \emph{bipartite} property --- \emph{concurrence} --- and two \emph{single-partite} properties --- \emph{coherence} and \emph{predictability}.
The \emph{concurrence} of the bipartite state is given by
\begin{equation}\label{scr C}
  \mathscr{C}(\Theta) = 2|ad-bc|.
\end{equation}
The quantity $\mathscr{C}$ provides a proper measure of bipartite entanglement. Particularly, $\mathscr{C}=0$ if and only if there is no entanglement (that is, $ad=bc$ if and only if $\ket{\Theta}$ can be cast as a product state), and $\mathscr{C}=1$ if and only if it has maximal entanglement.
The \emph{coherence} between the two orthogonal quibit states, $\ket{0}$ and $\ket{1}$, of the $k$-th particle ($k=1$ or $2$) is given by
\begin{equation}\label{scr V}
\mathscr{V}_1 = 2|ac^*+bd^*|,
\quad
\mathscr{V}_2 = 2|ab^*+cd^*|.
\end{equation}
The coherence $\mathscr{V}_k$ can be understood as visibility of the interference pattern, if one performs an interference experiment between the states $\ket{0}$ and $\ket{1}$ upon the $k$-th particle by introducing an adjustable relative phase shift between $\ket{0}$ and $\ket{1}$.
Finally, the \emph{predictability} of the $k$-th particle is given by
\begin{subequations}\label{scr P}
\begin{eqnarray}
\mathscr{P}_1 &=& \left|\left(|c|^2+|d|^2\right)-\left(|a|^2+|b|^2\right)\right|, \\
\mathscr{P}_2 &=& \left|\left(|b|^2+|d|^2\right)-\left(|a|^2+|c|^2\right)\right|,
\end{eqnarray}
\end{subequations}
which quantifies how well one can guess the outcome if one performs a qubit state measurement upon the $k$-th particle.
The condition $|a|^2+|b|^2+|c|^2+|d|^2=1$ leads to a ``triality'' complementarity relation:
\begin{equation}\label{triality relation}
  \mathscr{C}^2+\mathscr{V}_k^2+\mathscr{P}_k^2 = 1,
\end{equation}
which involves both single-partite and bipartite aspects.

Now, consider the subensembles associated with $0_d$ and $1_d$. In \figref{fig:quantum circuit}, the upper right grouped block and the lower right grouped block are independent of each other. Therefore, whether the measurement of $D_d$ is performed before or after that of $D_i$ makes no difference. In fact, we can pretend that the whole lower right block is performed before slice 3. In this case, the two-qubit state before the $D_d$ measurement is given by $\ket{\psi_5}$ in \eqref{psi5} with $\theta$ now replaced by $0$. That is, we have
\begin{eqnarray}\label{psi}
\ket{\psi} &=&
\frac{1}{\sqrt{2}}\left(\cos\frac{\phi+\phi'}{2}\ket{0}+\cos\frac{\phi-\phi'}{2}\ket{1}\right)
\ket{0} \nonumber\\
&& \quad \mbox{} + \frac{1}{\sqrt{2}}\left(\sin\frac{\phi+\phi'}{2}\ket{0}-\sin\frac{\phi-\phi'}{2}\ket{1}\right)
\ket{1}.
\end{eqnarray}
Then, after the measurement of $D_d$, the two-qubit state collapses into
\begin{subequations}\label{Theta}
\begin{eqnarray}
  \ket{\psi_{0_d}} &=&  \frac{1}{\sqrt{1+\cos\phi\cos\phi'}}\left(\cos\frac{\phi+\phi'}{2}\ket{0}+\cos\frac{\phi-\phi'}{2}\ket{1}\right)
\ket{0},\\
   \ket{\psi_{1_d}} &=&  \frac{1}{\sqrt{1-\cos\phi\cos\phi'}}\left(\sin\frac{\phi+\phi'}{2}\ket{0}-\sin\frac{\phi-\phi'}{2}\ket{1}\right)
\ket{1},
\end{eqnarray}
\end{subequations}
associated with the outcomes $0_d$ and $1_d$, respectively. The first qubit of the state \eqref{Theta} then enters the $P(\theta)$ gate and the $H$ gate and finally is detected by $D_i$. A moment of reflection on the physical meanings of the coherence $\mathscr{V}_{k=1}$ and predictability $\mathscr{P}_{k=1}$ leads to that $\mathscr{V}_{k=1}$ for the state \eqref{Theta} is identical to the visibility $\mathcal{V}_{0_d}$ or $\mathcal{V}_{1_d}$, and $\mathscr{P}_{k=1}$ is identical to the distinguishability $\left|\mathcal{D}_{0_d}\right|$ or $\left|\mathcal{D}_{1_d}\right|$. It is also easy to explicitly show that \eqref{scr V} and \eqref{scr P} for the state \eqref{Theta} yield the same results of $\mathcal{V}_{0_d}$, $\mathcal{V}_{1_d}$, $\left|\mathcal{D}_{0_d}\right|$, and $\left|\mathcal{D}_{1_d}\right|$ given by \eqref{V 0d}, \eqref{V 1d}, and \eqref{D 0d 1d}.
Furthermore, as the collapsed state \eqref{Theta} is a product state, the concurrence $\mathscr{C}$ vanishes. Therefore, the triality relation \eqref{triality relation} implies the saturated complementarity relation given by \eqref{V D duality sub}.

The triality relation can also be used to understand why the complementarity relation \eqref{V D duality} for the total ensemble is saturated when $\phi'=\pm\pi/2$. Again, we pretend that the $R_y(\phi')$ gate is performed before slice 3 in \figref{fig:quantum circuit}. The bipartite state before the measurements of $D_i$ and $D_d$ is again given by \eqref{psi}, for which \eqref{scr C}, \eqref{scr V} and \eqref{scr P} give rise to
\begin{subequations}\label{C V P}
\begin{eqnarray}
\label{scr C for psi}
  \mathscr{C} &=& |\sin\phi|, \\
\label{scr V for psi}
  \mathscr{V}_{k=1} &=& |\cos\phi|, \\
  \mathscr{P}_{k=1} &=& 0.
\end{eqnarray}
\end{subequations}
As expected, the coherence $\mathscr{V}_{k=1}$ given in \eqref{scr V for psi} is identical to the visibility $\mathcal{V}$ in \eqref{V} since they now have the same physical meaning.
In the special case of $\phi'=\pm\pi/2$, the gate $R_y(\phi')$ is equivalent to a $H$ gate (up to a sign difference and a swap between $\ket{0}$ and $\ket{1}$). For the open configuration in \figref{fig:quantum circuit} (i.e., the second $H$ gate is removed), the concurrence of the outcomes of $D_i$ and $D_d$ is the same as the concurrence $\mathscr{C}$ defined in \eqref{scr C}, because the first $H$ gate and the $R_y(\phi'=\pm\pi/2)$ gate are compensated with each other.\footnote{Also note that, in the open configuration, the phase gate $P(\theta)$ has no effect on the outcome of $D_i$.} It is also not difficult to argue that the concurrence of the outcomes of $D_i$ and $D_d$ is identical to the distinguishability $|\mathcal{D}|$ defined in \eqref{D}. Consequently, we have $\mathscr{C}=|\mathcal{D}|$. This is verified by the fact that $\mathscr{C}$ given in \eqref{scr C for psi} is identical to $|\mathcal{D}|$ in \eqref{D} if $\phi'=\pm\pi/2$. Therefore, the triality relation \eqref{triality relation} with \eqref{C V P} implies $\mathcal{V}^2+\mathcal{D}^2=1$ in \eqref{V D duality} in the case of $\phi'=\pm\pi/2$.

In the setting of \figref{fig:quantum circuit}, the which-way information is obtained through the entanglement between the two qubits. Therefore, it is natural that $|\mathcal{D}|\le\mathscr{C}$, and consequently the triality relation \eqref{triality relation} implies $\mathcal{V}^2+\mathcal{D}^2\le1$. Particularly, $|\mathcal{D}|$ yields the optimal value if the measurement of $D_d$ exploits the bipartite concurrence, that is, when $\phi'=\pm\pi/2$.

\section{Experiments on IBM Quantum}\label{sec:IBM Q}
In this section, we present the experimental results on the quantum circuit of superconducting qubits provided by IBM Quantum \cite{IBMQ}.
The results were obtained by the quantum computer \texttt{ibm\_auckland}. For comparison, more experimental results obtained by the noisier quantum computer \texttt{imbq\_toronto} are also provided in \appref{app:moredata}.
The details of the device layout and the system calibration data for both \texttt{ibm\_auckland} and \texttt{imbq\_toronto} can be found in \appref{app:calibration}.

In the circuit shown in \figref{fig:quantum circuit}, each time for the same setting of $\phi$, $\phi'$, and $\theta$, we perform 40000 shots for simulation and 5000 shots for experiment to accumulate measurement outcomes. For each shot, the measurements of $D_i$ and $D_d$ are performed in the computational (0/1) basis.
For a given pairs of $\phi$ and $\phi'$, we gradually vary the value of $\theta$ in the range from $0$ to $2\pi$ with the resolution of $0.04\pi$.
In the closed configuration (i.e., the dashed $H$ gate in \figref{fig:quantum circuit} is applied), varying $\theta$ with high resolution gives the interference pattern in response to $\theta$.
In the open configuration (i.e., the dashed $H$ gate in \figref{fig:quantum circuit} is removed), the measurement outcomes, theoretically, are independent of the value of $\theta$, but we nevertheless vary the values of $\theta$ to collect more data and average out any noise biased by the value of $\theta$.
Various probabilities such as $p(0_i)$, $p(0_i,0_d)$, etc.\ in the closed and open configurations are then counted as the relative frequencies of occurrence of the corresponding outcomes from the repetitive shots.

We also utilize the \emph{delay gate} in the Qiskit API to delay the measurement by $D_d$. The delay gate instructs the qubit to idle for a certain duration of delay, denoted as $t_{\text{delay}}$, which can be specified by the user in the unit of the sampling timestep $\text{dt}$ of the qubit drive channel \cite{Qiskit2021etal}. For the quantum computers used in this study, the timestep is $\text{dt}=0.22\,\text{ns}$.

\begin{figure}
	\includegraphics[width=0.9\textwidth]{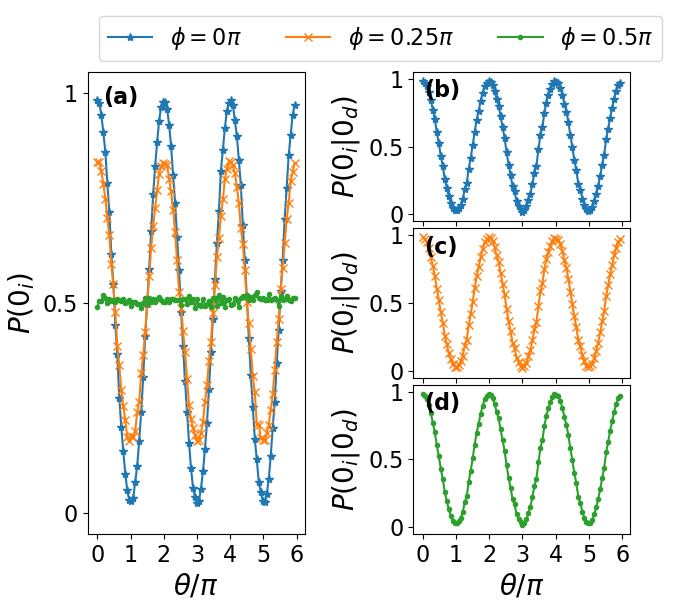}
	\caption{(a): The interference pattern of $p(0_i)$ for $\phi=0$, $0.25\pi$, and $0.5\pi$ performed on \texttt{ibm\_auckland}. (b--d): With $\phi'=0$, the interference pattern within the $0_d$ subensemble is depicted in terms of $p(0_i|0_d)$. The duration of delay is set to $t_{\text{delay}}=0\,\text{dt}$.}
	\label{fig:interf}
\end{figure}

Before studying the visibility-distinguishability complementarity relations, we demonstrate the behavior of delayed-choice quantum erasure in \figref{fig:interf}. We consider three cases of $\phi=0$, $0.25\pi$, and $0.5\pi$, which respectively correspond to zero, intermediate, and maximum entanglement between the two qubits. In the closed configuration, according to \eqref{p i total}, the probability $p(0_i)$ or $p(1_i)\equiv1-p(0_d)$ exhibits the interference pattern as modulated in response to the phase shift $\theta$ with the amplitude of modulation given by $|\cos\phi|/2$. (In the case of $\phi=0.5\pi$, the interference pattern is completely flatten.) The experimental results are shown in \figref{fig:interf}~(a), which closely agree with \eqref{p i total}.
Further, we take into account the outcomes of $D_d$ with $\phi'=0$. Within either of the two subensembles associated with $0_d$ and $1_d$, the interference pattern is maximally ``recovered'', i.e., $\mathcal{V}_{0_d}=\mathcal{V}_{1_d}=1$ according to \eqref{V 0d} and \eqref{V 1d}. This behavior of quantum erasure can be shown in terms of the conditional probabilities \eqref{p 0d} and \eqref{p 1d}. In \figref{fig:interf}~(b--d), in particular, we present the results of $p(0_i|0_d)$ for the three cases, for all of which the interference pattern is maximally restored with full contrast, regardless of $\phi$.

Next, we study the visibility-distinguishability relations. In particular, we choose $\phi'=0$, $0.25\pi$, and $0.5\pi$ for consideration. For each value of $\phi'$, we then consider $\phi=0,\ 0.1\pi,\ 0.2\pi,\ \dots,\ 2\pi$. In the closed configuration, for each pair of $\phi'$ and $\phi$, we obtain the interference pattern in response to $\theta$, and then the visibility is extracted from the maximum and minimum of the interference pattern according to the definitions in \eqref{V}, \eqref{V 0d}, \eqref{V 1d}, and \eqref{V avg}. For the same pair of $\phi'$ and $\phi$, we also perform the experiment in the open configuration, the distinguishability is then inferred from the correlation between the outcomes of $D_i$ and $D_d$ according to \eqref{p succ}, \eqref{D}, \eqref{D 0d 1d}, and \eqref{D avg}.
We first run the simulation on the \texttt{qasm} simulator without modeling any noise, and then perform the experiment on the \texttt{ibm\_auckland} quantum computer.
The same simulated and experimental data are analyzed in the following from three different perspectives as described in \secref{sec:V and D}.

\begin{figure}
	\includegraphics[width=\textwidth]{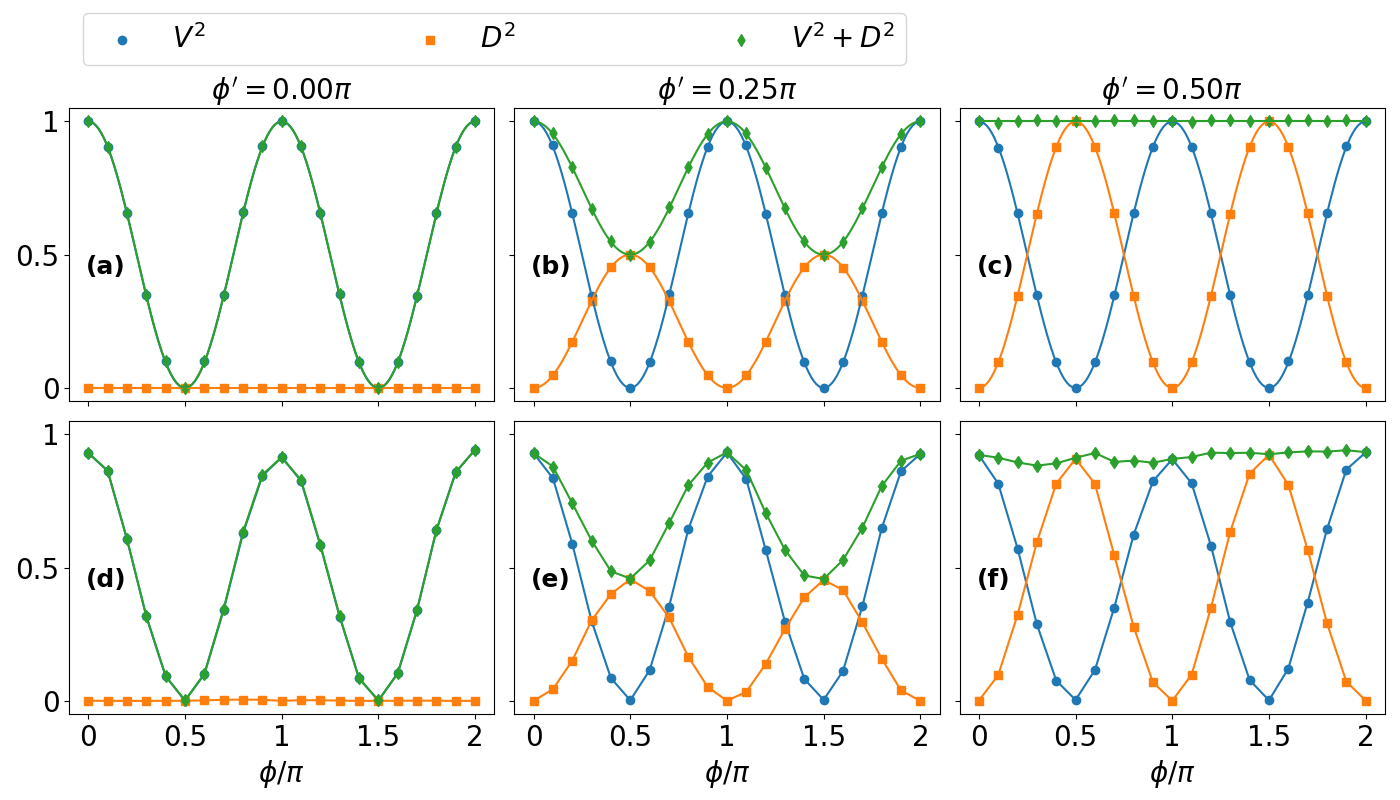}
	\caption{The interference visibility and path distinguishability in the total ensemble perspective of the simulation on \texttt{qasm} (a--c) and the experiment on \texttt{ibmq\_auckland} (d--f). The theoretical equations are depicted as the smooth curves in (a--c). From the left to right columns, $\phi^{\prime}=0$, $0.25\pi$, and $0.5\pi$, respectively.
For (d--f), the duration of delay is set to $t_\text{delay}=0\,\text{dt}$.}
	\label{fig:total}
\end{figure}

\begin{figure}
	\includegraphics[width=\textwidth]{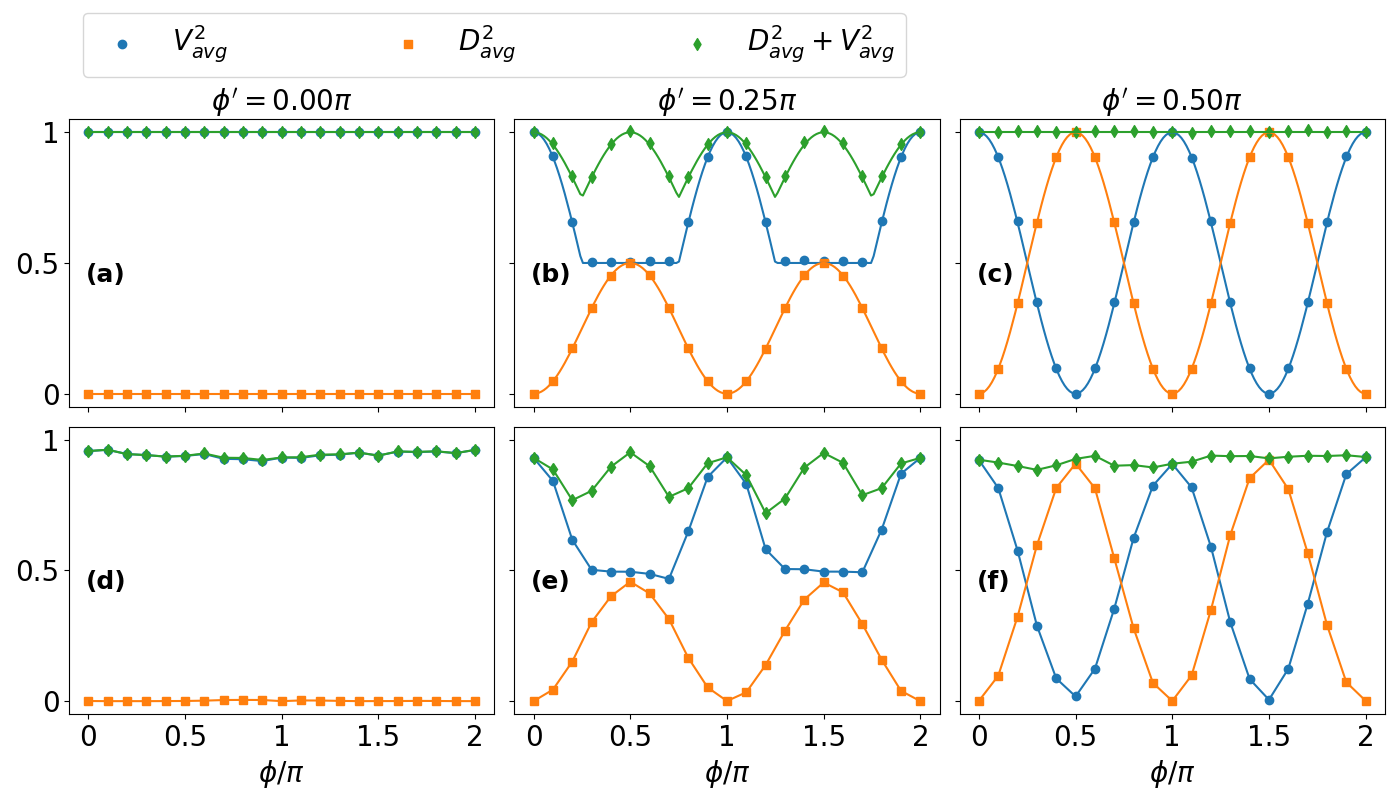}

    \caption{The interference visibility and path distinguishability in the average perspective of the simulation on \texttt{qasm} (a--c) and the experiment on \texttt{ibmq\_auckland} (d--f).}
	\label{fig:avg}
\end{figure}

\begin{figure}
	\includegraphics[width=\textwidth]{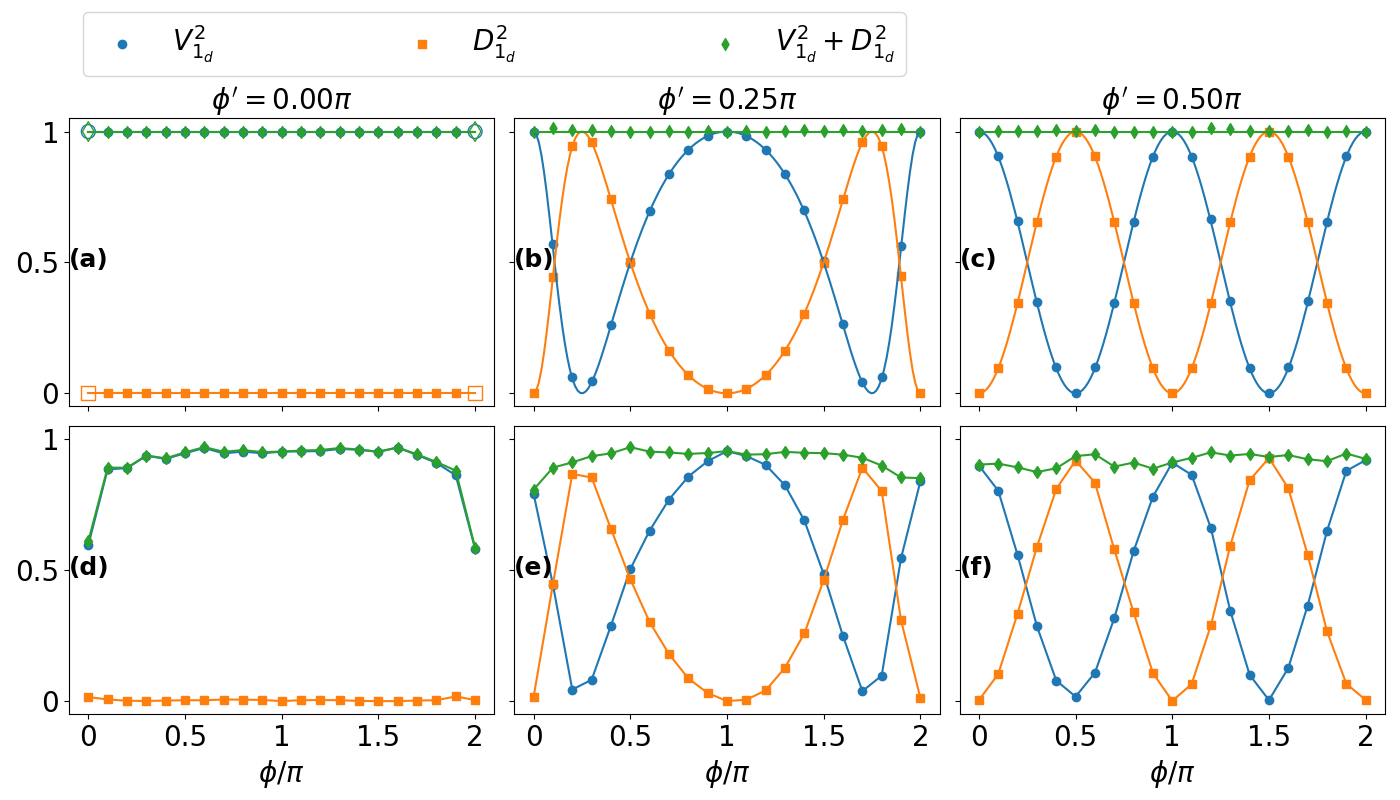}
	
	\caption{The interference visibility and path distinguishability in the $1_d$ subensemble perspective of the simulation on \texttt{qasm} (a--c) and the experiment on \texttt{ibmq\_auckland} (d--f).}
	\label{fig:sep}
\end{figure}

The results of interference visibility and path distinguishability in the total ensemble perspective are shown in \figref{fig:total}.
The top panels (a--c) are the simulated results obtained from the \texttt{qasm} simulator, which are in close agreement with the theoretical equations \eqref{V}, \eqref{D}, and \eqref{V D duality}.\footnote{The simulated data already agree closely with the theoretical ones when we perform 5000 shots for each setting. This ensures that the number of 5000 shots is large enough to average out probabilistic fluctuations for real experiments. In \figref{fig:total}--\figref{fig:sep}, we present the simulated data with 40000 shots for each setting. The deviation from the theory is almost inappreciable with the large number of 40000 shots.}
The bottom panels (d--f) are the experimental results obtained from \texttt{ibm\_auckland}, which agree well with the theoretical equations, in spite of slight deviations.

The results of interference visibility and path distinguishability in the average perspective are shown in \figref{fig:avg}.
The top panels (a--c) are the simulated results obtained from the \texttt{qasm} simulator, which are in close agreement with the theoretical equations \eqref{V avg}, \eqref{D avg}, and \eqref{V D duality avg}.
The bottom panels (d--f) are the experimental results obtained from \texttt{ibm\_auckland}, which agree well with the theoretical equations, in spite of slight deviations.
Comparing \figref{fig:total} with \figref{fig:avg}, we can affirm the effect of quantum erasure: the visibility $\mathcal{V}=|\cos\phi|$ in \figref{fig:total}, which is independent of $\phi'$, is fully or partially enhanced to $\mathcal{V}_\mathrm{avg}=\max(|\cos\phi|,|\cos\phi'|)$ in \figref{fig:avg}, depending on $\phi'$.

In both \figref{fig:total} and \figref{fig:avg}, the experimental results are slightly deviated from theoretical ones. Both visibility and distinguishability are slightly diminished from the theoretical values roughly by an overall diminishing factor.
However, the characteristic of the error is not universal but device dependent. For instance, the experimental results for the same circuit performed on another quantum computer, \texttt{ibmq\_toronto}, exhibit much more significant deviations in a more complicated way, which cannot be described by an overall diminishing factor (see the data in \appref{app:moredata})

The results of interference visibility and path distinguishability in the subensemble perspective are shown in \figref{fig:sep} (for the $1_d$ subensemble in particular).
The top panels (a--c) are the simulated results obtained from the \texttt{qasm} simulator, which are in close agreement with the theoretical equations \eqref{V 1d}, \eqref{D 0d 1d}, and \eqref{V D duality sub}.
The bottom panels (d--f) are the experimental results obtained from \texttt{ibm\_auckland}, which agree well with the theoretical equations, in spite of some deviations.
The deviation for $\mathcal{V}_{1_d}$ becomes rather significant when $\phi'=0$ and $\phi$ is close to $0$ or $2\pi$ in \figref{fig:sep}~(d).
As commented in \ftref{foot:special cases 2}, in the special case of $\phi'=0$ and $\phi=0\equiv2\pi$, the $1_d$ subensemble is supposed to be empty (i.e., $p(1_d)=0$), and thus $\mathcal{D}_{1_d}$ and $\mathcal{V}_{1_d}$ are both ill defined.\footnote{The simulated data indeed yield no $1_d$ events when $\phi'=0$ and $\phi=0\equiv2\pi$. In \figref{fig:sep}~(a), the points where $\mathcal{D}_{1_d}$ and $\mathcal{V}_{1_d}$ are ill defined are indicated by the hollow squares and hollow diamonds.}
In reality, the presence of noise allows $1_d$ events to occur (i.e., $p(1_d)\neq0)$. If the noise responsible for the occurrence of $1_d$ events is completely random, we should have $p(0_i|1_d)=p(1_i|1_d)=1/2$ both in the closed and open configurations, since the two qubits are completely unentangled for $\phi=0\equiv2\pi$. This renders $\mathcal{V}_{1_d}\approx0$ and $\mathcal{D}_{1_d}\approx0$ according to \eqref{V 1d} and \eqref{D 0d 1d}.
The experimental result of $\mathcal{V}_{1_d}$ in \figref{fig:sep}~(d) suffers significant deviation whenever $\phi$ is close to $0$ or $2\pi$, which, however, is quite different from $\mathcal{V}_{1_d}\approx0$. This suggests that the noise is not completely random but gives rise to unwanted entanglement between the two qubits.
In the following, we will demonstrate that this deviation can be attributed to the error of the CNOT gate.\footnote{By contrast, the same error does not lead to significant deviations in the total ensemble perspective and the average perspective as shown in \figref{fig:total} and \figref{fig:avg}.
The total visibility $\mathcal{V}$ defined in \eqref{V} does not involve $p(0_d)$ or $p(1_d)$ and thus is insusceptible to the errors on $p(0_d)$ and $p(1_d)$.
On the other hand, for $\mathcal{D}$ in \eqref{D} with \eqref{p succ}, $\mathcal{V}_\mathrm{avg}$ in \eqref{V avg}, an $\mathcal{D}_\mathrm{avg}$ in \eqref{D avg}, the errors on $p(1_i|1_d)$, $\mathcal{V}_{1_d}$, and $\mathcal{D}_{1_d}$ are greatly ``tamed'' through the multiplication by $p(1_d)$, which in the presence of noise remains close to $0$ when $\phi\approx0$ and $\phi\approx0\equiv2\pi$.
In fact, theoretically, $\mathcal{V}$, $\mathcal{D}$, $\mathcal{V}_\mathrm{avg}$, and $\mathcal{D}_\mathrm{avg}$ are all well defined even if $p(0_d)=0$ or $p(1_d)=0$ (recall \ftref{foot:special cases} and \ftref{foot:special cases 3}).}
The same error also gives rise to considerable deviation for $\mathcal{V}_{1_d}$ whenever $\phi$ is close to $0$ or $2\pi$ and the theoretical value of $p(1_d)$ remains small. In \figref{fig:sep}~(e), the deviation is noticeable when $\phi'=0.25\pi$ and $\phi$ is close to $0$ or $2\pi$. This is because, according to \eqref{p d total}, the theoretical value of $p(1_d)$ at the point of $\phi'=0.25\pi$ and $\phi=0\equiv2\pi$ is given by $p(1_d)=0.147$, which is still not large enough to be robust against the CNOT noise. On the other hand, in \figref{fig:sep}~(f), the deviation is much less noticeable, because theoretical value $p(1_d)=0.5$ at the point of $\phi'=0.5\pi$ and $\phi=0\equiv2\pi$ is quite large.

\begin{figure}
	\includegraphics[width=0.8\textwidth]{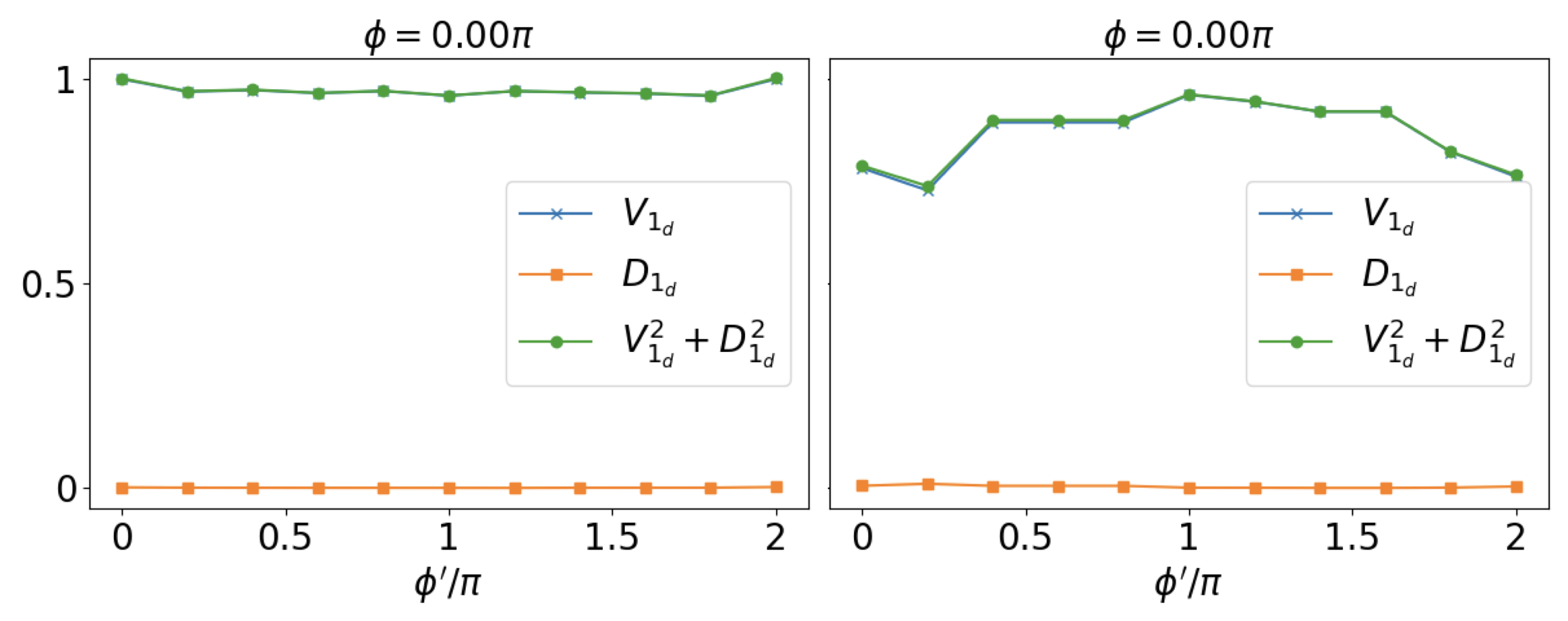}
	\caption{The experimental results of the $1_d$ subensemble for demonstrating the CNOT error by removing the $R_y(\phi)$ gate. (a) The CNOT gate is also removed. (b) The CNOT is present.}
\label{fig:cnoterr}
\end{figure}

In the case of $\phi'=0$ and $\phi=0\equiv2\pi$, theoretically, both the $R_{y}(\phi)$ gate and the CNOT gate in \figref{fig:quantum circuit} become superfluous, therefore producing the same result as if both gates were removed. However, in reality, the presence of the CNOT still make a difference due to its error.
To show that the CNOT gate error is responsible for the deviation mentioned above, we remove the $R_{y}(\phi)$ gate and compare the experiment results performed with and without the CNOT gate. The comparison is presented in \figref{fig:cnoterr}. Theoretically, the visibility $\mathcal{V}_{1_d}$ should not depend on $\phi'$ (the horizontal axis), since the two qubits are now completely unentangled. However, when the CNOT is present, as shown in \figref{fig:cnoterr}~(b), the visibility $\mathcal{V}_{1_d}$ is greatly dropped and becomes dependent on $\phi'$. This suggests that the CNOT error induces unnecessary entanglement between the two qubits.  

\begin{figure}
	\includegraphics[width=0.5\textwidth]{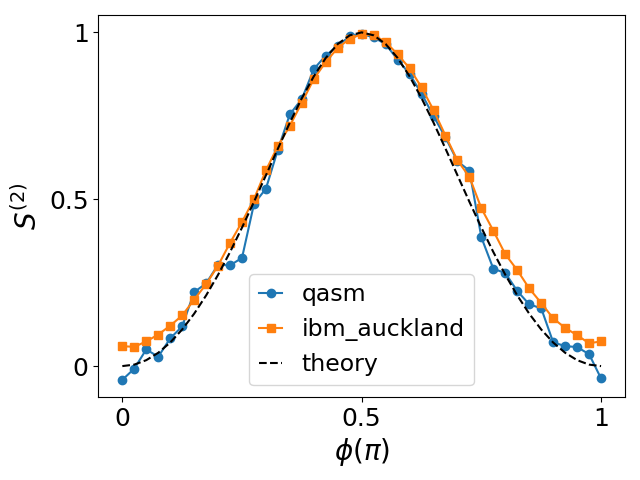}
	\caption{The second-order R\'{e}nyi enetropy for the quantum state at slice 2 in \figref{fig:quantum circuit}.}
	\label{fig:entropy}
\end{figure}

Moreover, we measure the second-order R\'{e}nyi entropy of the entangled state $|\psi_2\rangle$ at slice 2 in \figref{fig:quantum circuit} via the randomized measurement protocol \cite{Brydges2019}.
The second-order R\'{e}nyi entropy is defined as
\begin{eqnarray}
 	S^{(2)}=-\log_2{\gamma},
\end{eqnarray}
where $\gamma$ is the \emph{degree of purity}, the theoretical value of which is given by \eqref{purity}. The results are shown in \figref{fig:entropy}. For the results performed in \texttt{ibm\_auckland}, $S^{(2)}$ deviates from the theoretical value more significantly when $\phi$ approaches $0$ or $\pi$, whereby the ``i'' qubit is supposed to be a pure state instead of a mixed state. This suggests that the errors in the real device have the nature that prevents the qubit from staying in a pure state. 
The analysis of $S^{(2)}$ adds more evidence that the CNOT gate in the real device yields more entanglement than it is supposed to do.

Finally, it is noteworthy that, theoretically, the measurement of $D_d$ can be performed at any moment, even after the measurement of $D_i$, and still yields the same result. In a sense, the which-way information in the past can be \emph{retroactively} erased by the measurement of $D_d$ performed in a later time. In reality, however, the entanglement between the ``i'' and ``d'' qubits gradually corrupts over time due to decoherence. Therefore, in the closed configuration, if the measurement of $D_d$ is delayed longer, the interference pattern of the subensembles is enhanced by the quantum erasure effect to a lesser extent. Consequently, whereas $\mathcal{V}$ (which is independent of the measurement of $D_d$) remains the same as the nondelayed case, $\mathcal{V}_\mathrm{avg}$, and $\mathcal{V}_{0_d/1_d}$ diminish with the delay time.
For the same reason, in the open configuration, $\mathcal{D}$, $\mathcal{D}_\mathrm{avg}$, and $\mathcal{D}_{0_d/1_d}$ also diminish with the delay time.

\begin{figure}
	\includegraphics[width=0.9\textwidth]{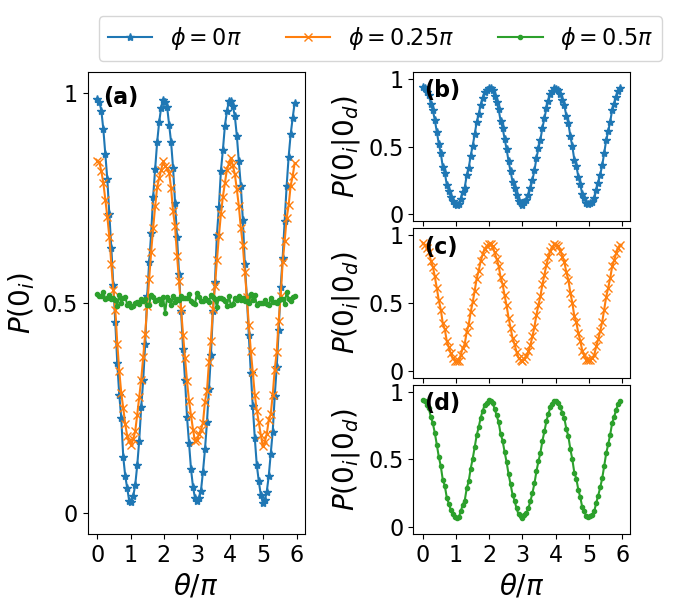}
	\caption{(a): The interference pattern of $p(0_i)$ for $\phi=0$, $0.25\pi$, and $0.5\pi$ performed on \texttt{ibm\_auckland}. (b--d): With $\phi'=0$, the interference pattern within the $0_d$ subensemble is depicted in terms of $p(0_i|0_d)$. The duration of delay is set to be $t_{\text{delay}}=5\times10^4\,\text{dt}=11\,\mu\text{s}$.}
	\label{fig:interf delay 50000}
\end{figure}

\begin{figure}
	\includegraphics[width=0.9\textwidth]{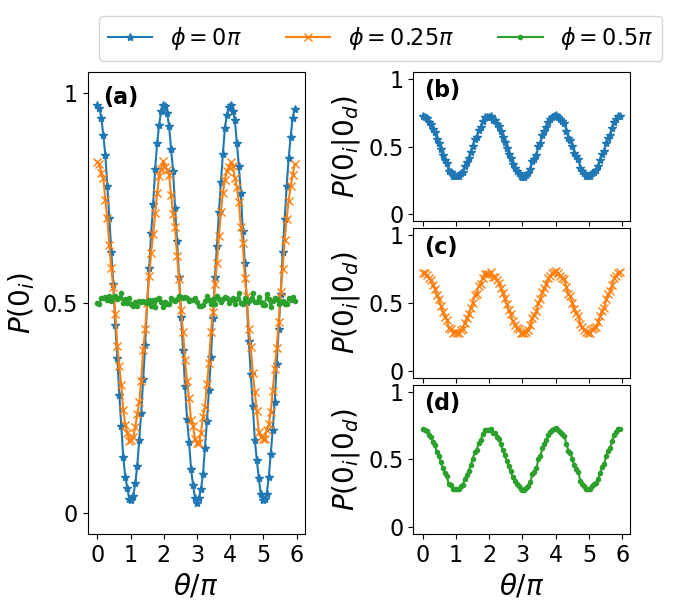}
	\caption{(a): The interference pattern of $p(0_i)$ for $\phi=0$, $0.25\pi$, and $0.5\pi$ performed on \texttt{ibm\_auckland}. (b--d): With $\phi'=0$, the interference pattern within the $0_d$ subensemble is depicted in terms of $p(0_i|0_d)$. The duration of delay is set to be $t_{\text{delay}}=5\times10^5\,\text{dt}=110\,\mu\text{s}$.}
	\label{fig:interf delay 500000}
\end{figure}

\begin{figure}
 	\includegraphics[width=\textwidth]{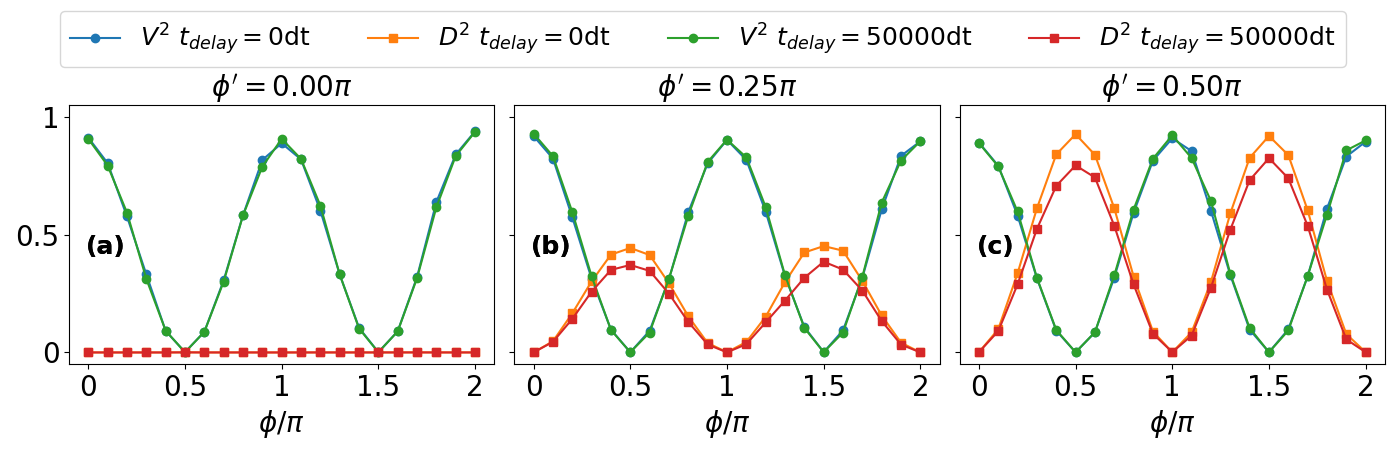}
 	\caption{The interference visibility and path distinguishability in the total ensemble perspective in comparison between the nondelayed and delayed cases. For the delayed measurement, $t_{\text{delay}}=5\times10^4\,\text{dt}=11\,\mu\text{s}$.}
 	\label{fig:total delay}
\end{figure}

\begin{figure}
 	\includegraphics[width=\textwidth]{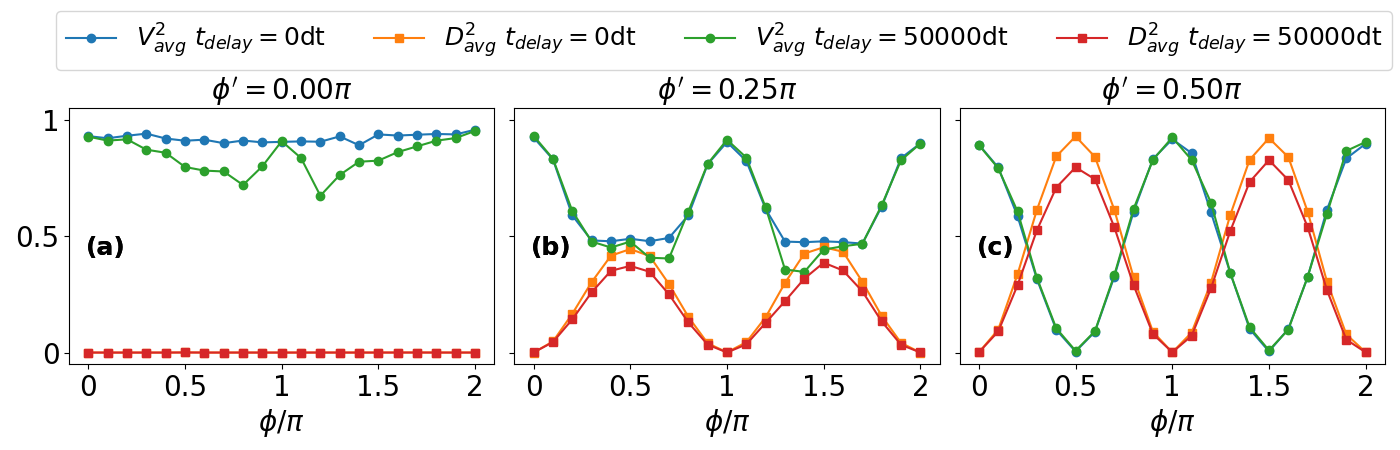}
 	\caption{The interference visibility and path distinguishability in the average perspective in comparison between the nondelayed and delayed cases. For the delayed measurement, $t_{\text{delay}}=5\times10^4\,\text{dt}=11\,\mu\text{s}$.
 	}
 	\label{fig:avg delay}
\end{figure}

\begin{figure}
 	\includegraphics[width=\textwidth]{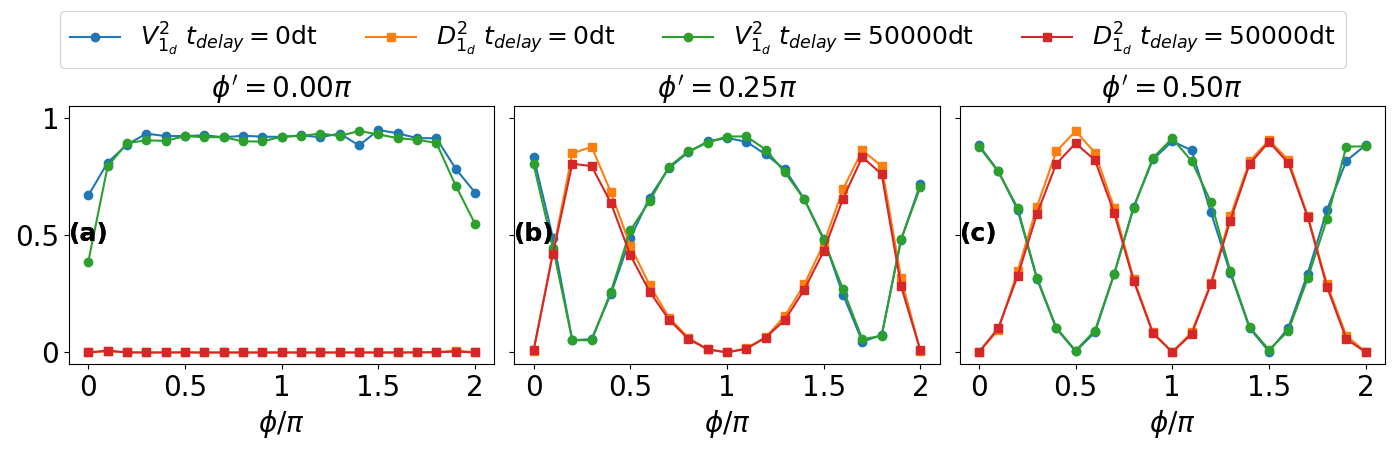}
 	\caption{The interference visibility and path distinguishability in the $1_d$ subensemble perspective in comparison between the nondelayed and delayed cases. For the delayed measurement, $t_{\text{delay}}=5\times10^4\,\text{dt}=11\,\mu\text{s}$.
 	}
 	\label{fig:sep delay}
\end{figure}

To affirm these phenomena, we apply the delay gate to delay the measurement of $D_d$ on \texttt{ibm\_auckland}.
The experimental results of the quantum erasure effect are presented in \figref{fig:interf delay 50000} for $t_{\text{delay}}=5\times 10^4\,\text{dt}=11\,\mu\text{s}$ and in \figref{fig:interf delay 500000} for $t_{\text{delay}}=5\times 10^5\,\text{dt}=110\,\mu\text{s}$. Compared with \figref{fig:interf}, the interference patterns in \figref{fig:interf delay 50000}~(b--c) and \figref{fig:interf delay 500000}~(b--c) are recovered to a lesser extent with contrast dropped slightly and considerably, respectively.
We also show the visibility and distinguishability from the total ensemble perspective, the average perspective, and the subensemble perspective in \figref{fig:total delay}, \figref{fig:avg delay}, and \figref{fig:sep delay}, respectively, for $t_{\text{delay}}=5\times 10^4\,\text{dt}$. The visibility and distinguishability still exhibit the trend expected by the theory. However, except for the total visibility $\mathcal{V}$, which remains almost unchanged, the visibility and distinguishability quantifiers $\mathcal{D}$, $\mathcal{V}_\mathrm{avg}$, $\mathcal{D}_\mathrm{avg}$, $\mathcal{V}_{0_d/1_d}$, and $\mathcal{D}_{0_d/1_d}$ in the delayed case all diminish to a certain degree compared to the nondelayed case.

\section{Summary and discussion}\label{sec:summary}
We propose a simple model of a delayed-choice quantum eraser as shown in \figref{fig:interferometer} using a Mach-Zehnder interferometer involving a pair of photons entangled in polarization. We then design a quantum circuit as shown in \figref{fig:quantum circuit} that emulates this quantum eraser model with the extension that the degree of entanglement between the two paired quantons is adjustable by turning $\phi$ in the $R_y(\phi)$ gate. This allows us to explore the intricate interplay between visibility, distinguishability, and entanglement in more depth.

Theoretically, we investigate complementarity relations between visibility and distinguishability from three different perspectives: (i) total ensemble perspective, (ii) subensemble perspective, and (iii) average perspective.
All complementarity relations obtained conform with the general duality relation \eqref{wave-particle duality}.

The complementarity relation for the total ensemble is given by \eqref{V D duality}, which can be understood as a special case in the entropic framework proposed in \cite{PhysRevA.93.062111}. The fact that \eqref{V D duality} saturates when the distinguishability $|\mathcal{D}|$ yields the optimal value at $\phi'=\pm\pi/2$ can be understood in terms of the triality relation \eqref{triality relation} for bipartite systems proposed in \cite{Jakob2010}. Because the which-way information is extracted through the bipartite entanglement, it is natural that $|\mathcal{D}|$ is optimized when the measurement of $D_d$ can fully exploit the bipartite concurrence, that is, when $\phi'=\pm\pi/2$.

Within either of the two subensembles associated with the readouts $0_d$ and $1_d$ of $D_d$ respectively, the complementarity relation is given by \eqref{V D duality sub}, which always saturates the general relation \eqnref{wave-particle duality}. This peculiar feature can be understood in view of of the triality relation \eqnref{triality relation}. Within the confines of each subensemble, the entangled bipartite state is collapsed into a product state and thus the concurrence $\mathscr{C}$ vanishes. As the coherence $\mathscr{V}_{k=1}$ is identical to the visibility $\mathcal{V}_{0_d}$ or $\mathcal{V}_{1_d}$ and the predictability $\mathscr{P}_{k=1}$ is identical to the distinguishability $\mathcal{V}_{0_d}$ or $\mathcal{V}_{1_d}$, the triality relation \eqnref{triality relation} with $\mathscr{C}=0$ implies the saturated relation \eqnref{V D duality sub}.

From the average perspective averaged over the two ensembles, the average visibility $\mathcal{V}_\mathrm{avg}$ defined in \eqnref{V avg} and the average distinguishability $\mathcal{D}_\mathrm{avg}$ defined in \eqnref{D avg} satisfy the complementarity relation \eqnref{V D duality avg}. In view of the entropic framework in \cite{PhysRevA.93.062111}, this is a special case of the generic wave-particle duality \eqnref{V D duality erasure} from the viewpoint of ``quantum erasure''. Whereas the average distinguishability $\mathcal{D}_\mathrm{avg}=-\sin\phi\sin\phi'$ is identical to the total-ensemble distinguishability $\mathcal{D}$, the average visibility $\mathcal{V}_\mathrm{avg}=\max\left(|\cos\phi|, |\cos\phi'|\right)$ is greater than or equal to the total-ensemble visibility $\mathcal{V}=|\cos\phi|$. The interference visibility is said to be enhanced by the ``quantum erasure'' scenario, but nevertheless the sum of the squares of visibility and distinguishability is still bound from above by unity.

We then perform simulations and experiments on IBM Quantum.
The simulated and experimental results are analyzed in the three perspectives, and the results are all in good agreement with the theoretical calculations, as shown in \figref{fig:total}, \figref{fig:avg}, and \figref{fig:sep}.
The only considerable deviation occurs in some extreme situations when $\phi$ is close to $0$ or $2\pi$ from the subensemble perspective. Our analysis suggests that this deviation is due to the error of the CNOT gate that gives rise to unwanted entanglement between the two qubits. This error, on the other hand, does not lead to significant deviations in the total ensemble perspective and the average perspective.

Furthermore, we apply the delay gate to delay the measurement of $D_d$. Compared with the results without any delay, the interference pattern within either of the two subensembles associated with the readouts $0_d$ and $1_d$ of $D_d$ still exhibits the quantum erasure effect, but the interference pattern is recovered to a lesser extent as shown in \figref{fig:interf delay 50000} and \figref{fig:interf delay 500000} due to decoherence over time, which corrupts the entanglement between the two qubits. Because of the same effect of decoherence, as shown in \figref{fig:total delay}, \figref{fig:avg delay}, and \figref{fig:sep delay}, the visibility and distinguishability quantifiers $\mathcal{D}$, $\mathcal{V}_\mathrm{avg}$, $\mathcal{D}_\mathrm{avg}$, $\mathcal{V}_{0_d/1_d}$, and $\mathcal{D}_{0_d/1_d}$ all diminish to a certain degree compared to the nondelayed case.

Finally, it is noteworthy that the various quantifiers of visibility and distinguishabilty considered in this paper from different perspectives are all empirically measurable. The fact that visibility is measured in the closed configuration whereas distinguishability is measured in the open configuration is in line with Bohr's complementarity principle in the sense that the measurement of a certain physical property inherently excludes the measurement of its complementary counterpart.
However, if one considers the idea of quantum-controlled experiments as proposed in \cite{PhysRevLett.107.230406}, it seems possible to have complementary behaviors simultaneously in a single experimental setup. This suggests that the complementarity principle might need to be revised in order to incorporate the ``morphing behavior''.
On the other hand, the work of \cite{dieguez2022experimental} defines the elements of physical reality of being a wave or a particle for the morphing states in terms of contextual realism, and shows that Bohr's original complementarity principle remains valid in a quantum-controlled experiment where physical reality of the morphing state is defined at each instant of time. The elements of physical reality defined in \cite{dieguez2022experimental} bear sound ontological meanings, but they seems to be counterfactual in nature (i.e., not empirically measurable). In our opinion, whether and how the complementarity principle should be revised remains a question open to further scrutiny. In any case, it will yield considerable insight into the issues of physical reality of complementary phenomena if one can investigate the complementarity relations in a delayed-choice experiment in more depth with the extension taking quantum-controlled devices into account.

\newpage

\begin{acknowledgments}
This work was supported in part by the National Science and Technology Council, Taiwan under the Grants 111-2112-M-004-008, 111-2119-M-007-009, 110-2112-M-110-015, 111-2112-M-110-013, 111-2119-M-002-012, and 112-2119-M-002-017. The availability of accessing the IBM Quantum cloud services via the IBM Q Hub at National Taiwan University is gratefully acknowledged.
\end{acknowledgments}

\appendix

\section{An elementary account of \eqref{V D duality sub}}\label{app:comp}
It is notable that the complementarity relation \eqref{V D duality sub} from the separate subensemble perspective is saturated to unity. Its meaning and significance can be understood in view of the triality relation \eqref{triality relation} as discussed in \secref{sec:triality relation}. Here, we present a simple account for \eqref{V D duality sub}.

First, we express the state at slice 5 in \figref{fig:quantum circuit} in the generic form
\begin{eqnarray}
	\ket{\psi_5}=a\ket{00}+b\ket{01}+c\ket{10}+d\ket{11}.
\end{eqnarray}
The state at slice 6 is then given by
\begin{eqnarray}
\ket{\psi_6} &=&H\otimes\mathbbm{1} \ket{\psi_5}\nonumber\\
&=&\frac{1}{\sqrt{2}}\big((a+b)\ket{00}+(a-b)\ket{01}+(c+d)\ket{10}+(c-d)\ket{11}\big).
\end{eqnarray}
The contrast for the $0_d$ subensemble defined in \eqref{C 0d} is computed from $\ket{\psi_6}$ as
\begin{eqnarray}
	\mathcal{C}_{0_d}&=& p(0_i|0_d) - p(1_i|0_d) \nonumber\\
	&=&\frac{|a+b|^2-|a-b|^2}{|a+b|^2+|a-b|^2} = \frac{a^*b+ab^*}{|a|^2+|b|^2}.	
\end{eqnarray}
Because the phase gate $P(\theta)$ acting on the interference qubit, we have $a^*b+ab^*=2|a||b|\cos\theta$.
Consequently, the visibility $\mathcal{V}_{0_d}$ defined in \eqref{V 0d} is given by
\begin{eqnarray}
	\mathcal{V}_{0_d}=\frac{2|a||b|}{|a|^2+|b|^2}.
\end{eqnarray}
Meanwhile, the distinguishability for the $0_d$ subensemble defined in \eqref{D 0d 1d} is computed from $\ket{\psi_5}$ as
\begin{eqnarray}
\mathcal{D}_{0_d}=2p(0_i|0_d)-1
=\frac{|a|^2-|b|^2}{|a|^2+|b|^2}.
\end{eqnarray}
It follows that $\mathcal{D}_{0_d}^2+\mathcal{V}_{0_d}^2=1$. In the subensemble perspective, the complimentary relation always saturates.

\section{System calibration data}\label{app:calibration}
\begin{figure}
		\includegraphics[width=0.65\textwidth]{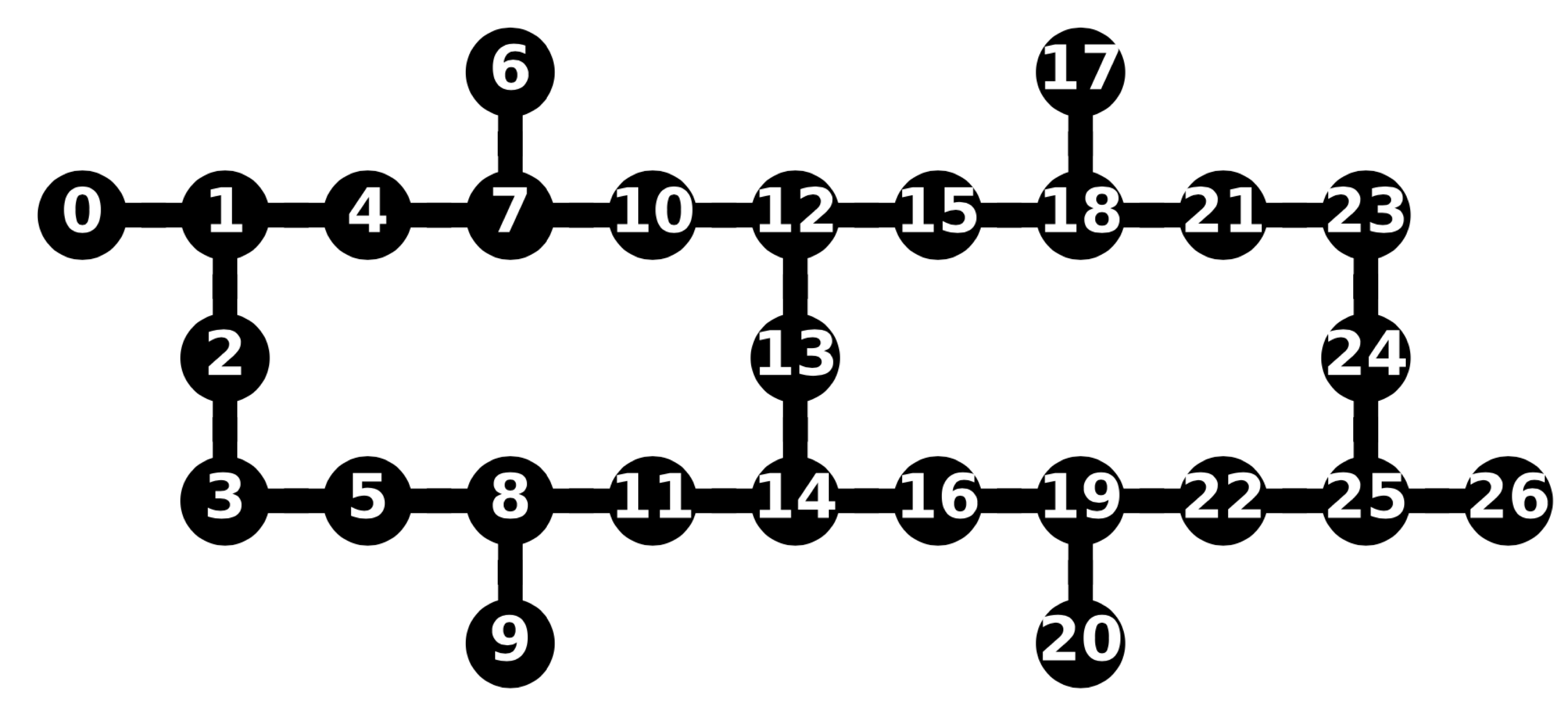}
	\caption{The qubit layout of the IBM Q machines \texttt{ibm\_auckland} and \texttt{ibmq\_toronto}. The numbers index the physical qubits and the edges denote the connections that allow CNOT operations.
	}
	\label{fig:layout}
\end{figure}

The experimental results presented in the main text are performed on \texttt{ibm\_auckland}, and more experimental results presented in \appref{app:moredata} are performed on \texttt{ibmq\_toronto}. Both \texttt{ibm\_auckland} and \texttt{ibmq\_toronto} share the same qubit layout as shown in \figref{fig:layout}.\footnote{This picture of the layout is obtained from the Qiskit API \cite{Qiskit2021etal}.}

We select three different pairs for our ``i'' and ``d'' qubits: (i) physical qubits 4 and 7 on \texttt{ibm\_auckland}, (ii) physical qubits 1 and 4 on \texttt{ibm\_auckland}, and (iii) physical qubits 1 and 4 on \texttt{ibmq\_toronto}.
The calibration data of these pairs were retrieved directly from the IBM Q website on the same days of the experiments and are reported in Table~\ref{tab:calib}.
Pair (i) is used for \figref{fig:total}--\figref{fig:sep}. Pair (ii) is used for \figref{fig:interf} and \figref{fig:cnoterr}--\figref{fig:sep delay}.
Pair (iii) is used for the figures in \appref{app:moredata}.

\begin{table}[]
	\begin{tabular}{|c|c|c|c|c|c|}
		\hline
		pair & machines  & \stackanchor{physical}{qubits} & $T_1\ (\mu\text{s})$ & $T_2\ (\mu\text{s})$ & \stackanchor{CNOT}{error rate}       \\ \hline 		\multirow{2}{*}{(i)}                                                                & \multirow{2}{*}{\texttt{ibm\_auckland}} & 4                                                                              & 277.64                            & 359.54                            & \multirow{2}{*}{$3.653\times 10^{-3}$}  \\ \cline{3-5}
		&                                & 7                                                                              & 221.56                            & 202.29                            &                                        \\ \hline
		\multirow{2}{*}{(ii)}                                                                & \multirow{2}{*}{\texttt{ibm\_auckland}} & 1                                                                              & 200.88                            & 203.58                            & \multirow{2}{*}{$6.071\times 10^{-3}$} \\ \cline{3-5}
		&                                & 4                                                                              & 250.98                            & 246                               &                                        \\ \hline
		\multirow{2}{*}{(iii)}                                                                & \multirow{2}{*}{\texttt{ibmq\_toronto}} & 1                                                                              & 136.34                            & 113.5                             & \multirow{2}{*}{$1.17\times 10^{-2}$}  \\ \cline{3-5}
		&                                & 4                                                                              & 117.56                            & 151.47                            &                                        \\ \hline
	\end{tabular}
\caption{The calibration data of the selected physical qubits for the pairs we have chosen.
The data were retrieved directly from the IBM Quantum website on the same days of the experiments.
$T_1$ is the longitudinal (thermal) relaxation time, and $T_2$ is the transverse (dephasing) relaxation time.}
\label{tab:calib}
\end{table}


\section{More experimental data}\label{app:moredata}
In order to know more about the nature of noises in the real devices of IBM Quantum, in addition to experiments performed on \texttt{ibm\_auckland}, we also perform experiments on \texttt{ibmq\_toronto} for comparison.

\begin{figure}
	\includegraphics[width=\textwidth]{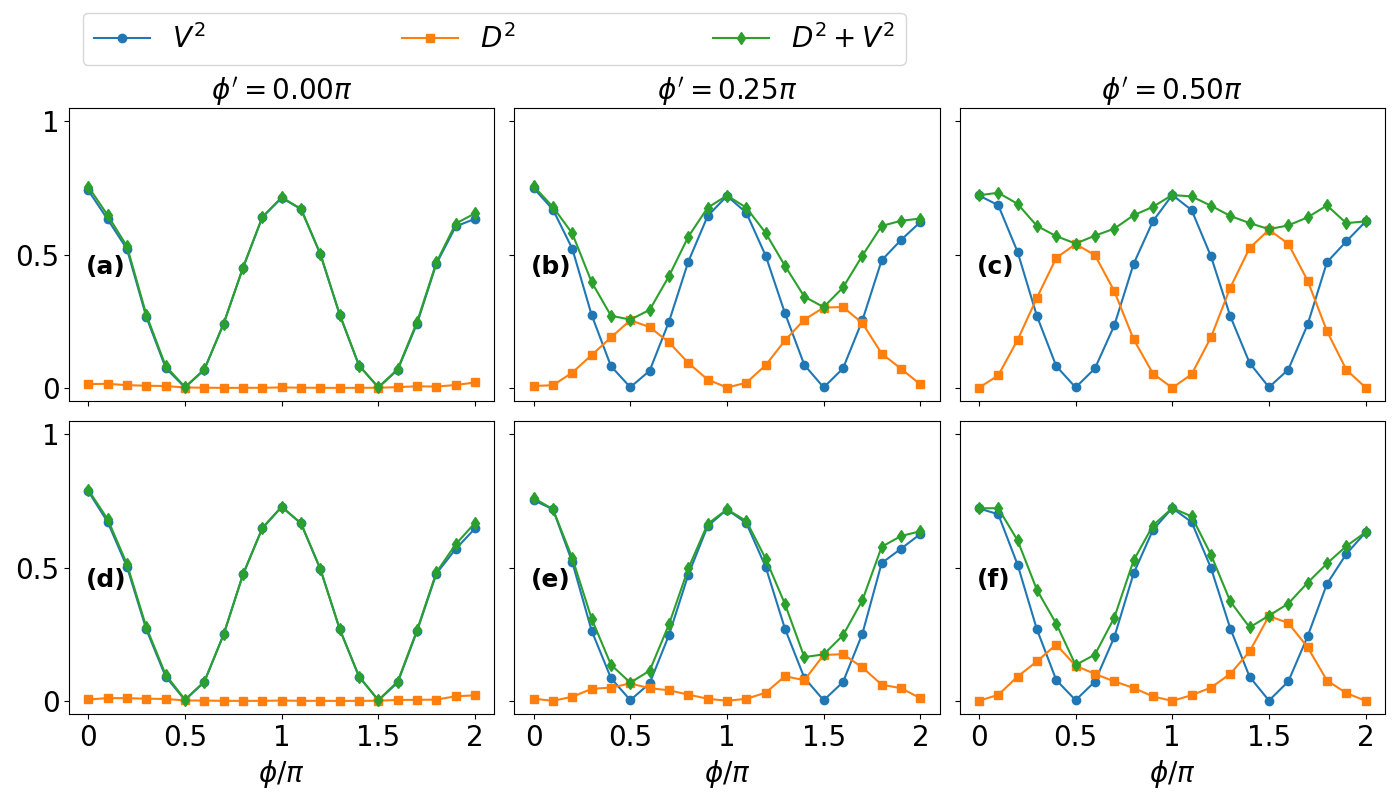}
	\caption{The interference visibility and path distinguishability in the total ensemble perspective on \texttt{ibmq\_toronto} with $t_\text{delay}=0$ (a--c) and $t_\text{delay}=100\,\text{dt}$ (d--f). From the left to right colomns, $\phi^{\prime}=0$, $0.25\pi$, and $0.5\pi$.
	}
	\label{fig:apptotal}
\end{figure}

\begin{figure}
	\includegraphics[width=\textwidth]{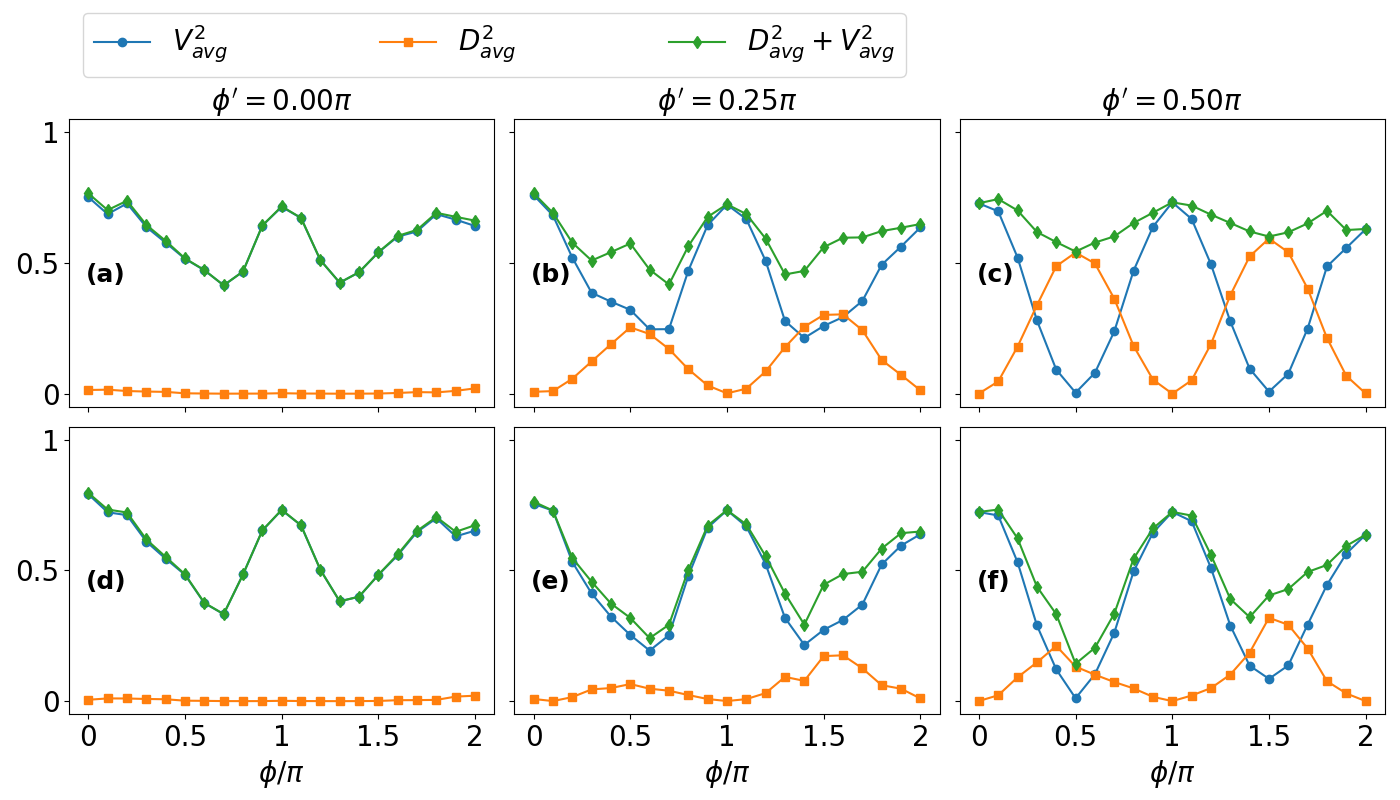}
	\caption{The interference visibility and path distinguishability in the average perspective on \texttt{ibmq\_toronto} with $t_\text{delay}=0$ (a--c) and $t_\text{delay}=100\,\text{dt}$ (d--f).
	}
	\label{fig:appavg}
\end{figure}

\begin{figure}
	\includegraphics[width=\textwidth]{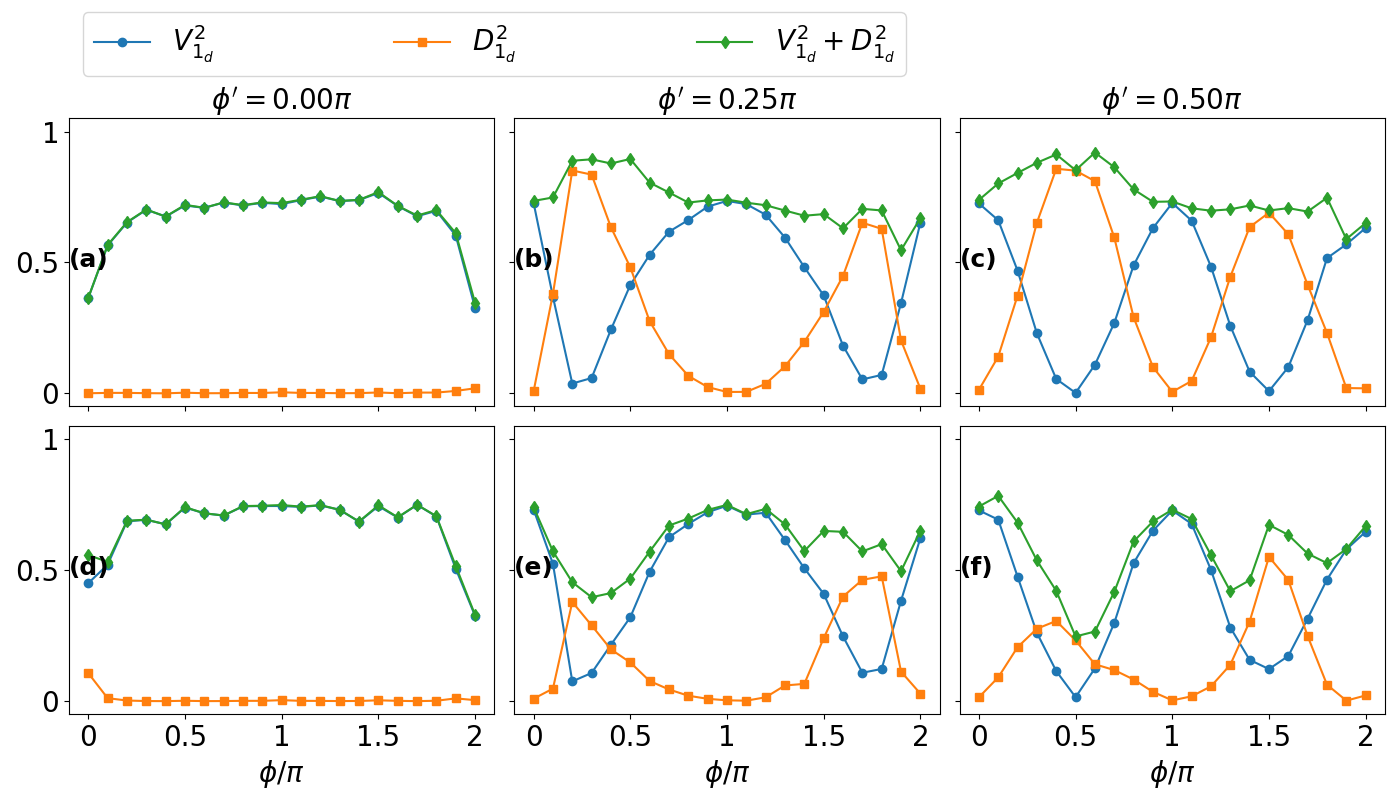}
	\caption{The interference visibility and path distinguishability in the $1_d$ perspective on \texttt{ibmq\_toronto} with $t_\text{delay}=0$ (a--c) and $t_\text{delay}=100\,\text{dt}$ (d--f).
	}
	\label{fig:appsep}
\end{figure}

The experimental results on \texttt{ibmq\_toronto} from the total ensemble perspective, the average perspective, and the subensemble perspective are presented in \figref{fig:apptotal}, \figref{fig:appavg}, and \figref{fig:appsep}, respectively. The top panels (a--c) show the results with the delay time $t_{\text{delay}}=0\,\text{dt}$ in the delay gate, while the bottom panels (d--f) show the results with $t_{\text{delay}}=100\,\text{dt}$.
Compared with the experimental results performed on \texttt{ibm\_auckland} as presented in the main text, the visibility and distinguishability are much noisier.
Moreover, the time delay has stronger effect on \texttt{ibmq\_toronto} than on \texttt{ibm\_auckland}: the effect is appreciable with $t_{\text{delay}}\sim100\,\text{dt}$ on the former, but it is not appreciable until $t_{\text{delay}}\sim10^5\,\text{dt}$ on the latter.
The fact that \texttt{ibmq\_toronto} is much noisier than \texttt{ibm\_auckland} is in agreement with the fact that the former has shorter relaxation times and a larger CNOT error rate, according to the daily calibration data provided by IBM Quantum as shown in Table~\ref{tab:calib}.


%


%

\end{document}